\documentclass[aps,prd,twocolumn,superscriptaddress,showpacs,preprintnumbers,floatfix,letterpaper,10pt]{revtex4}
\usepackage{graphicx,color,here}
\newcommand{\be}{\begin{equation}}
\newcommand{\ee}{\end{equation}}
\newcommand{\bea}{\begin{eqnarray}}
\newcommand{\eea}{\end{eqnarray}}

\newcommand{\bfk}{\mbox{\boldmath $k$}}
\def\bkt{\bfk_\perp}
\def\kt{k_\perp}
\newcommand{\bfp}{\mbox{\boldmath $p$}}
\def\bpo{{\bfp}_{\perp 1}}
\def\bpt{{\bfp}_{\perp 2}}
\def\bpp{\bfp_\perp}
\newcommand{\bfq}{\mbox{\boldmath $q$}}
\newcommand{\bfP}{\mbox{\boldmath $P$}}
\newcommand{\bfS}{\mbox{\boldmath $S$}}
\newcommand{\bfs}{\mbox{\boldmath $s$}}

\def\ppo{p_{\perp 1}}
\def\ppt{p_{\perp 2}}
\def\pp{p_\perp}
\newcommand{\pup}{p^\uparrow}
\newcommand{\qup}{q^\uparrow}
\newcommand{\qdown}{q^\downarrow}

\newcommand{\la}{\lambda}

\newcommand{\ua}{\uparrow}
\newcommand{\da}{\downarrow}
\def\xb{x_{_{\!B}}}
\def\avk{\langle k_\perp ^2\rangle}
\def\avp{\langle p_\perp ^2\rangle}
\def\avk{\langle k_\perp ^2\rangle}
\def\avp{\langle p_\perp ^2\rangle}
\def\avPT{\langle P_T^2\rangle}

\def\T{_{_T}}
\def\C{_{_C}}

\def\lsim{\mathrel{\rlap{\lower4pt\hbox{\hskip1pt$\sim$}}\raise1pt\hbox{$<$}}}
\def\gsim{\mathrel{\rlap{\lower4pt\hbox{\hskip1pt$\sim$}}\raise1pt\hbox{$>$}}}
\def\nostrocostruttino#1\over#2{\mathrel{\mathop{\kern 0pt \rlap
{\hbox{$#1$}}} \hbox{\kern-.135em $#2$}}}

\begin{document}
%%%%%%%%%%%%%%%%%%%%%%%%%%%%%%%%%%%%%%%%%%%%%%%%%%%%%%%%%%%%%%%%%%%%%%%%%%%%%%
%\preprint{}
\title{Transversity and Collins functions from SIDIS and $e^+e^-$ data}

\author{M.~Anselmino}
\affiliation{Dipartimento di Fisica Teorica, Universit\`a di Torino and \\
          INFN, Sezione di Torino, Via P. Giuria 1, I-10125 Torino, Italy}
\author{M.~Boglione}
\affiliation{Dipartimento di Fisica Teorica, Universit\`a di Torino and \\
          INFN, Sezione di Torino, Via P. Giuria 1, I-10125 Torino, Italy}
\author{U.~D'Alesio}
\affiliation{Dipartimento di Fisica, Universit\`a di Cagliari and
          INFN, Sezione di Cagliari,\\
          C.P. 170, I-09042 Monserrato (CA), Italy}
\author{A.~Kotzinian}
\affiliation{Yerevan Physics Institute, 375036 Yerevan, Armenia, \\
          JINR, 141980 Dubna, Russia, and \\
          INFN, Sezione di Torino, Via P. Giuria 1, I-10125 Torino, Italy
}

\author{F.~Murgia}
\affiliation{Dipartimento di Fisica, Universit\`a di Cagliari and
          INFN, Sezione di Cagliari,\\
          C.P. 170, I-09042 Monserrato (CA), Italy}
\author{A.~Prokudin}
\affiliation{Dipartimento di Fisica Teorica, Universit\`a di Torino and \\
          INFN, Sezione di Torino, Via P. Giuria 1, I-10125 Torino, Italy}
\author{C.~T\"{u}rk}
\affiliation{Dipartimento di Fisica Teorica, Universit\`a di Torino and \\
          INFN, Sezione di Torino, Via P. Giuria 1, I-10125 Torino, Italy}
%
%\date{December 31, 2006}

\begin{abstract}
A global analysis of the experimental data on azimuthal asymmetries in
semi-inclusive deep inelastic scattering (SIDIS),
from the HERMES and COMPASS Collaborations, and in $e^+e^- \to h_1 h_2 X$
processes, from the Belle Collaboration, is performed. It results in the
extraction of the Collins fragmentation function and,
\textit{for the first time}, of the transversity distribution function for
$u$ and $d$ quarks. These turn out to have opposite signs and to be sizably
smaller than their positivity bounds. Predictions for the azimuthal asymmetry
$A_{UT}^{\sin(\phi_S + \phi_h)}$, as will soon be measured at JLab and
COMPASS operating on a transversely polarized proton target, are then
presented.
\end{abstract}


\maketitle

\section{\label{Intro} Introduction}

The transversity distribution function, usually denoted as
$h_{1q}(x, Q^2)$ or $\Delta_T q(x, Q^2)$, together with the unpolarized
distribution functions $q(x, Q^2)$ and the helicity distributions
$\Delta q(x, Q^2)$, contains basic and necessary information for a full
understanding of the quark structure, in the collinear, $\bfk_\perp$
integrated configuration, of a polarized nucleon.
The distribution of transversely polarized quarks in a transversely
polarized nucleon, $\Delta_T q(x, Q^2)$, is so far unmeasured. The reason is
that, being related to the expectation value of a chiral-odd quark operator,
it appears in physical processes which require a quark helicity flip: this
cannot be achieved in the usual inclusive DIS, due to the helicity conservation
of perturbative QED and QCD processes.

The problem of measuring the transversity distribution has been largely
discussed in the literature~\cite{Barone:2001sp}. The most promising approach
is considered the double transverse spin asymmetry $A_{TT}$ in Drell-Yan
processes in $p \bar p$ interactions at a squared c.m. energy of the order of
200 GeV$^2$, which has been proposed by the PAX Collaboration
\cite{Barone:2005pu,Anselmino:2004ki,Efremov:2004qs,Pasquini:2006iv}. However, this requires
the availability of polarized antiprotons, which is an interesting, but
formidable task in itself. Meanwhile, the most accessible channel, which
involves the convolution of the transversity distribution with the Collins
fragmentation function \cite{Collins:1992kk}, is the azimuthal asymmetry
$A_{UT}^{\sin(\phi_S + \phi_h)}$ in SIDIS processes, namely
$\ell \, \pup \to \ell \, \pi \, X$. This is the strategy being pursued by
HERMES, COMPASS and JLab Collaborations.

A crucial improvement, towards the success of this strategy, has been recently
achieved thanks to the independent measurement of the Collins function (or
rather, of the convolution of two Collins functions), in
$e^+e^- \to h_1 h_2 \, X$ unpolarized processes by Belle Collaboration at
KEK \cite{Abe:2005zx}. By combining the SIDIS experimental data from
HERMES \cite{Airapetian:2004tw,HERMES:proceedings} and COMPASS \cite{Ageev:2006da}, with the
Belle data, we have, for the first time, a large enough set of data points
as to attempt a global fit which involves, as unknown functions, both the
transversity distributions and the Collins fragmentation functions of
$u$ and $d$ quarks.

In Section~\ref{sidis} we briefly remind the basic formalism involved in the
description of the SIDIS asymmetry $A_{UT}^{\sin(\phi_S+\phi_h)}$, and in
Section~\ref{eplus-eminus} we develop, in somewhat greater detail, a similar formalism
for the azimuthal correlations, involving two Collins functions, measured
by Belle in $e^+e^- \to h_1 h_2 \, X$ processes. In Section~\ref{fit} 
we perform
a global fit of HERMES \cite{Airapetian:2004tw,HERMES:proceedings}, COMPASS \cite{Ageev:2006da}
and Belle \cite{Abe:2005zx} data, in order to extract \textit{simultaneously}
the Collins fragmentation function $\Delta^N D_{\pi/\qup}(z, \pp)$ and the
transversity distribution function $\Delta_T q(x)$ for $q = u,d$.
We then use, in Section~\ref{pred}, the transversity distributions and the Collins
functions so determined, to give predictions for forthcoming experiments at
JLab and CERN-COMPASS. Comments and conclusions are gathered in 
Section~\ref{comm}.

\section{\label{sidis} Transversity and Collins functions from SIDIS processes}

\begin{figure}[t]
\begin{center}
\scalebox{0.30}{\input{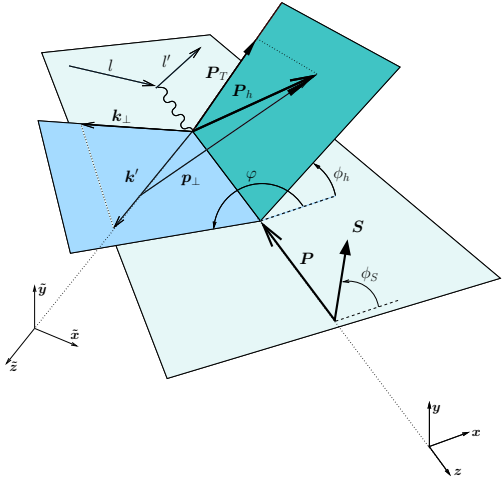}}
\caption{\small Three dimensional kinematics of the SIDIS process, according
to Trento conventions \cite{Bacchetta:2004jz}. The photon and the proton
collide along the $\hat z$-axis, while the leptonic plane defines the
$\hat{xz}$ plane. The fragmenting quark and the final hadron $h$ are emitted
at azimuthal angles $\varphi$ and $\phi_h$, and the proton transverse spin
direction is identified by $\phi_S$.}
\label{fig:planessidis}
\end{center}
\end{figure}

The exact kinematics for SIDIS $\ell \, p \to \ell \, h \, X$
processes in the $\gamma ^* -p$ c.m. frame, including all intrinsic motions,
was extensively discussed in Ref.~\cite{Anselmino:2005nn}, and is schematically
represented in Fig.~\ref{fig:planessidis}. We take the virtual photon and the
proton colliding along the $\hat z$-axis with momenta $\bfq$ and $\bfP$
respectively, and the leptonic plane to coincide with the $\hat{xz}$ plane.
We work in the kinematic regime in which
$P_T \simeq \Lambda_{\rm QCD} \simeq \kt$, where $\kt$ is the magnitude of the
intrinsic transverse momentum $\bkt$ of the initial quark with respect
to the parent proton and $P_T = |\bfP_T|$ is the magnitude of the final hadron
transverse momentum. We neglect second order corrections in the $\kt/Q$
expansion: in this approximation, the transverse momentum $\bfp _\perp$ of the
observed hadron $h$ with respect to the direction of the fragmenting quark
is related to $\bfk _\perp$ and $\bfP _T$ by the simple expression
\(\bpp = \bfP _T - z \bfk_\perp\); in addition, the lightcone momentum
fractions $x$ and $z$ coincide with the usual measurable SIDIS variables,
$z = z_h = (P \cdot P_h)/(P \cdot q)$ and  $x = \xb = Q^2/(2P \cdot q)$.
In this region factorization holds \cite{Ji:2004xq,Ji:2004wu}, leading order
$\ell \, q \to \ell \, q$ elementary processes are dominating and the soft
$P_T$ of the detected hadron is mainly originating from intrinsic motions.
%
%%%%%%%%%%%%%%%%%%%%%%%%%%%%%%%%%%%%%%%%%%%%%%%%%%%%%%%%%%%%%%%%%%%%%%%%%%%
%%%%%%%%%%%%%%%%%%%%%%%%%%%%%%%%%%%%%%%%%%%%%%%%%%%%%%%%%%%%%%%%%%%%%%%%%%%
%

The transverse single spin asymmetry (SSA) for this process is defined as
\be
A_{_{UT}}=
\frac{\displaystyle
      d^6\sigma^{\ell p^\uparrow \to \ell^\prime h X} \,-\,
      d^6\sigma^{\ell p^\downarrow \to \ell^\prime h X}}
     {\displaystyle
      d^6\sigma^{\ell p^\uparrow \to \ell^\prime h X} \,+\,
      d^6\sigma^{\ell p^\downarrow \to \ell^\prime h X}} \equiv
\frac{d\sigma^\uparrow - d\sigma^\downarrow}
     {d\sigma^\uparrow + d\sigma^\downarrow}\,,
\label{sidis-asym}
\ee
where $d^6\sigma^{\ell p^{\ua,\da} \to \ell h X} \equiv
d\sigma^{\uparrow,\downarrow}$ is a short hand notation for
$(d^6\sigma^{\ell p^{\ua,\da} \to \ell h X })/(d\xb \,
dy \, dz_h \, d^2 {\bfP}_T \, d\phi_S)$. It will often happen, in comparing
with data or giving measurable predictions, that the numerator and denominator
of Eq.~(\ref{sidis-asym}) will be integrated over some of the variables,
according to the kinematical coverage of the experiments.
$\ua$ and $\da$ refer, respectively, to polarization vectors
$\bfS$ and $-\bfS$, see Fig.~\ref{fig:planessidis}.
A full study of Eq.~(\ref{sidis-asym}), with all contributions at all
orders in $\kt/Q$, will be presented in a forthcoming paper
\cite{sidis-general}.

We consider here, at ${\cal O}(\kt/Q)$, the $\sin(\phi_S +\phi_h)$
weighted asymmetry,
\be
A^{\sin (\phi_S + \phi_h)}_{_{UT}} = \label{defsin-asym}
2 \, \frac{\int d\phi_S \, d\phi_h \,
[d\sigma^\uparrow - d\sigma^\downarrow] \, \sin(\phi_S +\phi_h)}
{\int d\phi_S \, d\phi_h \,
[d\sigma^\uparrow + d\sigma^\downarrow]}\,,
\ee
measured by the HERMES \cite{Airapetian:2004tw,HERMES:proceedings} and COMPASS
\cite{Ageev:2006da} Collaborations.
This asymmetry singles out the spin dependent part of the fragmentation
function of a transversely polarized quark with spin polarization $\hat{\bfs}$ and
three-momentum $\bfp_q$:
\bea
D_{h/q,s}(z,\bpp) &=& D_{h/q}(z,\pp) \nonumber \\ &+& \frac{1}{2} \, \Delta^N
D_{h/\qup}(z,\pp)\, \hat{\bfs}\cdot(\hat{\bfp}_q \times \hat{\bfp}_\perp) \,,
\;\;\;
\label{Collins-gen}
\eea
resulting in
\begin{widetext}
\be
A^{\sin (\phi_S+\phi_h)}_{_{UT}} = \label{sin-asym}
\frac{\displaystyle  \sum_q e_q^2  \! \! \int \! \!{d\phi_S \, d\phi_h \, d^2
\bfk _\perp}\,\Delta _T q (x,\kt) \,
\frac{d (\Delta {\hat \sigma})}{dy}\,
\Delta^N D_{h/q^\ua}(z,\pp) \sin(\phi_S + \varphi +\phi_q^h)
\sin(\phi_S +\phi_h) } {\displaystyle \sum_q e_q^2 \, \int {d\phi_S
\,d\phi_h \, d^2 \bfk _\perp}\; f_{q/p}(x,k _\perp) \;
\frac{d\hat\sigma}{dy}\; \; D_{h/q}(z,p_\perp) } \, \cdot
\ee
\end{widetext}
In the above equation $\Delta_Tq(x,\kt)$ is the unintegrated transversity
distribution,
\be
\Delta_Tq(x) \equiv h_{1q}(x) = \int d^2\bkt \, \Delta_Tq(x, \kt) \,,\label{h1}
\ee
while $\Delta^N D_{h/q^\ua}(z,\pp)$ is the Collins function, often denoted
as \cite{Bacchetta:2004jz}:
\be
\Delta^N D_{h/q^\ua}(z,\pp) = \frac{2\pp}{z m_h} \;
H_1^{\perp q}(z,\pp) \;. \label{colmul}
\ee
$d{\hat \sigma}/dy$ is the planar unpolarized elementary cross section
\be
\frac{d{\hat \sigma}}{dy}=\frac{2\pi\alpha^2}{sxy^2}\,[1+(1-y)^2]\;,
\ee
and
\be
\frac{d(\Delta {\hat \sigma})}{dy} =
\frac{d{\hat \sigma}^{\ell \qup \to \ell \qup}}{dy} -
\frac{d{\hat \sigma}^{\ell \qup \to \ell \qdown}}{dy} =
\frac{4\pi\alpha^2}{sxy^2}\,(1-y)\, \cdot
\ee
The $\sin(\phi_S + \varphi +\phi_q^h)$ azimuthal dependence in
Eq.~(\ref{sin-asym}) arises from the combination of the phase factors in the
transversity distribution function, in the non-planar $\ell\,q \to \ell\,q$
elementary scattering amplitudes, and in the Collins fragmentation function;
$\phi_S$ and $\varphi$ identify the directions of the proton spin $\bfS$
and of the quark intrinsic transverse momentum $\bfk_\perp$, see
Fig.~\ref{fig:planessidis}; $\phi_q^h$ is the azimuthal angle of
the final hadron $h$, as defined in the fragmenting
quark helicity frame. Neglecting ${\cal O}(\kt ^2/Q^2)$ terms, one finds
\bea
\cos\phi_q^h &=&
\frac{P_T}{\pp}\,\cos(\phi_h-\varphi) -z\,\frac{\kt}{\pp}\,,
\nonumber \\
\sin\phi_q^h &=& \frac{P_T}{\pp}\,\sin(\phi_h-\varphi)\,.\label{phases}
\eea
A full study of Eq.~(\ref{defsin-asym}), taking into account intrinsic
motions with all contributions at all orders, following the general approach
of Ref. \cite{Anselmino:2005sh}, will be presented in a forthcoming paper
\cite{sidis-general}. Here, in agreement with all papers on the Collins
effect in SIDIS so far appeared in the literature, we work at
${\cal O}(\kt /Q)$ and use Eqs. (\ref{sin-asym}) and (\ref{phases}).

$f_{q/p}(x,k _\perp)$ is the unpolarized transverse momentum dependent (TMD)
distribution function of a quark $q$ inside the parent proton $p$, while
$D_{h/q}(z,p_\perp)$ is the unpolarized TMD
fragmentation function of quark $q$ into the final hadron $h$.
We assume the $k _\perp$ and $p_\perp$ dependences of these functions to
be factorized in a Gaussian form, suitable to describe non-perturbative
effects at small $P_T$ values and simple enough to allow analytical
integration over the intrinsic transverse momenta:
\bea
&& f_{q/p}(x,\kt)=
   f_{q/p}(x)\;\frac{e^{-{\kt^2}/\avk}}{\pi\avk}\;,\\
&& D_{h/q}(z,\pp)=D_{h/q}(z)\;\frac{e^{-\pp^2/\avp}}{\pi\avp}\;,
\label{unpfrag}
\eea
where $f_{q/p}(x)$ and $D_{h/q}(z)$ are the usual integrated parton
distribution and fragmentation functions, available in the literature; in
particular we refer to Refs.~\cite{Gluck:1998xa,Gluck:2000dy}
and \cite{Kretzer:2000yf}. The QCD induced $Q^2$ dependence of these functions
is also taken into account, although we do not indicate it explicitly.
Finally, the average values of $\kt^2$ and $\pp^2$ are taken from
Ref.~\cite{Anselmino:2005nn}, where they were obtained by fitting the
azimuthal dependence of SIDIS unpolarized cross section:
\be
\langle \kt^2 \rangle = 0.25 \> \textrm{GeV}^2 \,,\quad\quad
\langle \pp^2 \rangle = 0.20 \> \textrm{GeV}^2 \>. \label{ktpp}
\ee
Notice that such values are assumed to be constant and flavor independent.

The transversity distributions and the Collins functions are unknown.
We choose the following simple parameterization
\be
\Delta_T q(x, \kt) =
\frac{1}{2} \, {\cal N}^{\T}_q(x)\,
\left[f_{q/p}(x)+\Delta q(x) \right] \;
\frac{e^{-{\kt^2}/{\avk\T}}}{\pi \avk \T} \label{tr-funct} \,, 
\ee
\be 
\Delta^N D_{h/q^\uparrow}(z,\pp) = 2\,{\cal N}^{\C}_q(z)\;
D_{h/q}(z)\;h(\pp)\,\frac{e^{-\pp^2/{\avp}}}{\pi \avp}\;,
\label{coll-funct}
\ee
with
\bea
&& {\cal N}^{\T}_q(x)= N^{\T}_q \,x^{\alpha} (1-x)^{\beta} \,
\frac{(\alpha + \beta)^{(\alpha +\beta)}} {\alpha^{\alpha} \beta^{\beta}}\,,
\label{NT}\\
&&{\cal N}^{\C}_q(z)= N^{\C}_q \, z^{\gamma} (1-z)^{\delta} \,
\frac{(\gamma + \delta)^{(\gamma +\delta)}}
{\gamma^{\gamma} \delta^{\delta}}\,, \label{NC}\\
&&
h(\pp)=\sqrt{2e}\,\frac{p_\perp}{M}\,e^{-{p_\perp^2}/{M^2}}\,,\label{h-funct}
\eea
and $|N^{\T}_q|, \> |N^{\C}_q| \le 1$. In general $\avk \T \neq \avk$,
but from our fits we learn that present experimental data are insensitive to
such a difference, therefore we simply assume $\avk \T = \avk$. Also, in this
first simultaneous extraction of the transversity and Collins functions, we
let the coefficients $N^{\T}_q$ and $N^{\C}_q$ to be flavor dependent
($q = u,d)$, while all the exponents $\alpha, \beta, \gamma, \delta$ and the
dimensional parameter $M$ are taken to be flavor independent.

Notice that our parameterizations are devised in such a way that the
transversity distribution function automatically obeys the Soffer
bound \cite{Soffer:1994ww}
\be
|\Delta_T q(x)| \le \frac{1}{2}
\left[ f_{q/p}(x) + \Delta q(x)\right ]\,,
\label{soffer}
\ee
and the Collins function satisfies the positivity bound
\be
|\Delta^N D_{h/q^\uparrow}(z, p_\perp)| \le 2 D_{h/q}(z, p_\perp) \,,
\label{bound}
\ee
since ${\cal N}^{\T}_q(x)$, ${\cal N}^{\C}_q(z)$ and $h(\pp)$ are normalized to
be smaller than $1$ in size for any value of $x$, $z$ and $\pp$ respectively.

By insertion of the above expressions into Eq.~(\ref{sin-asym}), we obtain,
in agreement with Refs.~\cite{Kotzinian:1994dv,Mulders:1995dh},
\begin{widetext}
\be
A^{\sin (\phi_S+\phi_h)}_{_{UT}} =
\frac{\displaystyle  \frac{P_T}{M}\,\frac{1-y}{s x y^2}\,
\sqrt{2e} \, \frac{\avp ^2 \C}{\avp}
\, \frac{e^{-P_T^2/\avPT \C}}{\avPT ^2 \C} \sum_q e_q^2 \;
 {\cal N}^{\T}_q(x)
\left[f_{q/p}(x)+\Delta q(x) \right]\;
{\cal N}^{\C}_q(z)\,
D_{h/q}(z)}
{ \displaystyle \frac{e^{-P_T^2/\avPT}}{\avPT} \,
\frac{[1+(1-y)^2]}{s x y^2}\,
 \sum_q e_q^2 \, f_{q/p}(x)\; D_{h/q}(z)}\;,
\label{sin-asym-final}
\ee
\end{widetext}
where
\be
\avp \C= \frac{M^2 \avp}{M^2 +\avp}\,, \quad\quad
 \avPT=\avp+z^2\avk \,,\quad\quad
\avPT \C=\avp \C+z^2\avk\,.
\ee
%\end{widetext}
%\bea
%&& \avp \C= \frac{M^2 \avp}{M^2 +\avp} \nonumber  \,,\\
%&& \avPT=\avp+z^2\avk \,, \frac{}{}\\
%&& \avPT \C=\avp \C+z^2\avk  \nonumber \,.
%\eea
Eq.~(\ref{sin-asym-final}) expresses $A^{\sin(\phi_S+\phi_h)}_{_{UT}}$ in
terms of the parameters  $\alpha, \beta, \gamma, \delta, N^{\T}_q, N^{\C}_q$
and $M$. In Section~\ref{fit} we shall fix them by performing a best fit of the
measurements of HERMES, COMPASS and Belle Collaborations. Actually, we shall
consider, following the experimental data, $A^{\sin (\phi_S+\phi_h)}_{_{UT}}$
as a function of one variable at a time, by properly integrating the numerator
and denominator of Eq.~(\ref{sin-asym-final}): the integration over $x$ and
$z$ gives the $P_T$ distribution of $A^{\sin (\phi_S+\phi_h)}_{_{UT}}$,
whereas the integrations over $P_T$ and $z$ or $P_T$ and $x$, yield the $x$
and  $z$ distributions.
Notice that, with our approximations, $x=\xb$ and $z=z_h$.

\section{\label{eplus-eminus}Collins functions from $e^+e^-$ processes}

\begin{figure}[t]
\begin{center}
\scalebox{0.36}{\input{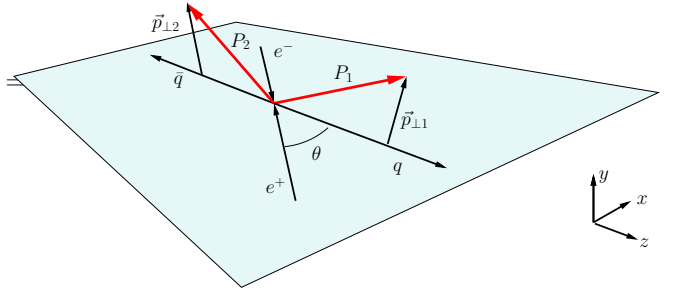}}
\caption{\small Three dimensional kinematics of the $e^+e^-\to h_1 h_2\,X$
process, in the $q \, \bar q$ c.m. frame. In this configuration the
reconstructed thrust axis identifies the $\hat z$-direction, the
lepton-quark scattering plane defines the $\hat{xz}$ plane.}
\label{fig:epluseminus}
\end{center}
\end{figure}

The kinematics corresponding to the $e^+e^-\to h_1 h_2 \, X$ process is
schematically represented in Fig.~\ref{fig:epluseminus}: the two detected
hadrons $h_1$ and $h_2$ are the fragmentation products of a quark
and an antiquark originating from $e^+e^-$ collisions. We choose
the reference frame so that the $e^+e^-\to q \, \bar q$ scattering
occurs in the $\hat{xz}$ plane, with the back-to-back quark and antiquark
moving along the $\hat{z}$-axis. This choice requires, experimentally,
the reconstruction of the jet thrust axis, but it involves a very simple
kinematics and a direct contribution of the Collins functions, as we shall
see. A different choice, originally suggested in the literature
\cite{Boer:1997mf}, is discussed at the end of this Section. In the
configuration of Fig.~\ref{fig:epluseminus}, the four-momenta of the
$e^+, e^- \,(k^+, k^-)$ and of the $q, \bar q \> (q_1, q_2)$ are

\bea
q_1 \!\!\!&=&\!\!\! \frac{\sqrt{s}}{2}(1,0,0,1)\,, \quad\quad
%\nonumber \\ && 
q_2 = \frac{\sqrt{s}}{2}(1,0,0,-1)\,,\\  
k^-\!\!\!\!\!&=&\!\!\! \frac{\sqrt{s}}{2}(1,\!\!-\!\sin\theta,0,\cos\theta) \,,
\quad
%\nonumber\\&& 
k^+\!\! =\!\frac{\sqrt{s}}{2}(1,\sin\theta,0,\!\!-\!\cos\theta) \nonumber \>.
\eea
The final hadrons $h_1$ and $h_2$ carry lightcone momentum fractions
$z_1$ and $z_2$ and have intrinsic transverse momenta $\bpo$
and $\bpt$ with respect to the direction of fragmenting quarks,
\bea
\bpo &=&\ppo(\cos\varphi_1,\sin\varphi_1,0) \,,\nonumber \\
\bpt &=&\ppt(\cos\varphi_2,\sin\varphi_2,0)\,,
\eea
so that their four-momenta can be expressed as
\bea
P_1&=&\left( z_1\,\frac{\sqrt{s}}{2}+\frac{\ppo^2}{2z_1\sqrt{s}}\,,
        \;\ppo \, \cos\varphi_1\,, \;\ppo \, \sin\varphi_1\,, \right. 
\nonumber \\
   & & \left.z_1\,\frac{\sqrt{s}}{2}-\frac{\ppo^2}{2z_1\sqrt{s}} \right) ,\\
P_2&=&\left( z_2\,\frac{\sqrt{s}}{2}+\frac{\ppt^2}{2z_2\sqrt{s}}\;,
        \;\ppt \, \cos\varphi_2\;, \;\ppt \, \sin\varphi_2\;,\right.\nonumber \\
   & &\left.\!-z_2\,\frac{\sqrt{s}}{2}+\frac{\ppt^2}{2z_2\sqrt{s}} \right) .
\eea
At large c.m. energies and not too small values of $z$, one can neglect
second order corrections in the $\pp/(z\sqrt{s})$ expansion, to work with the
much simpler kinematics:
\bea
\!\!P_1\!\!&=&\!\!\!\left(\! z_1\,\frac{\sqrt{s}}{2},
        \;\ppo \cos\varphi_1, \;\ppo \, \sin\varphi_1,
        \;z_1\,\frac{\sqrt{s}}{2} \right),\;\; \\
\!\!P_2\!\!&=&\!\!\!\left(\! z_2\,\frac{\sqrt{s}}{2},
        \;\ppt \, \cos\varphi_2, \;\ppt \, \sin\varphi_2,
        \;-z_2\,\frac{\sqrt{s}}{2} \right) .\;\;
\eea
Notice also that in this limit the lightcone momentum fractions $z$ coincide
with the observable energy fractions $z_h$,
\be
z_h = 2 E_h/\sqrt{s} = z + \frac{\pp^2}{zs} \simeq z \>.
\ee
%
%%%%%%%%%%%%%%%%%%%%%%%%%%%%%%%%%%%%%%%%%%%%%%%%%%%%%%%%%%%%%%%%%%%%%%%%%%%
%%%%%%%%%%%%%%%%%%%%%%%%%%%%%%%%%%%%%%%%%%%%%%%%%%%%%%%%%%%%%%%%%%%%%%%%%%%
%
The cross section corresponding to this process, with unpolarized leptons, 
can be written as
\bea
\label{general-belle}
&&\frac{d\sigma ^{e^+e^- \to h_1 h_2 X}}
{dz_1\,dz_2\,d^2\bpo\,d^2\bpt\,d\cos\theta}=\\ 
&& \hspace*{1.6cm}
\frac{3}{32\pi s} \sum _q \frac{1}{4}\sum_{\{\lambda\}} 
\hat M _{\la_q \la_{\bar q}; \la_+ \la_-}  \,
\hat M ^* _{\la^\prime_q \la^\prime_{\bar q}; \la_+ \la_-} 
\nonumber \\ && \hspace*{1.6cm}
\hspace*{1.4cm}\times \; D^{h_1/q}_{\la _q \la ^\prime_q}(z_1,\bpo) 
\, D^{h_2/\bar q}_{\la_{\bar q} \la ^\prime_{\bar q}}(z_2,\bpt)\,, \nonumber
\eea
where 
$\hat M _{\la_q \la_{\bar q}; \la_+ \la_-}$ are the helicity amplitudes 
corresponding to the elementary scattering 
$e^+(\la _+)e^-(\la _-) \to q(\la _q)\bar q(\la_{\bar q})$, 
$q = u, \bar u, d, \bar d, s, \bar s$ (neglecting heavy flavors) and 
$\sum _{\{\lambda\}}$ indicates a sum over all helicity indices. 
%Notice that the initial leptons are not polarized, 
%therefore $\rho_{\la_+\la^+\prime_+} \, \rho_{\la_-\la^\prime_-} = 1/4$.
In this case there are only two non-zero, independent amplitudes:
\bea
\hat M _{+-;+-} = \hat M _{-+;-+} = e^2 e_q (1+\cos\theta) \,,\nonumber \\
\hat M _{-+;+-} = \hat M _{+-;-+} = e^2 e_q (1-\cos\theta) \,.
\label{hel-amp}
\eea
The functions
$D^{h_1/q}_{\la_q \la^\prime_q}(z_1,\bpo)$ and 
$D^{h_2/\bar q}_{\la_{\bar q} \la^\prime_{\bar q}}(z_2,\bpt)$ are the 
probability densities which describe the fragmentation of  
quarks and antiquarks into the physical hadrons $h_1$ and $h_2$ respectively 
(see Section II.C of Ref.~\cite{Anselmino:2005sh} for detailed explanations). 
In particular, the diagonal elements $D^{h/q}_{++}(z,\bpp)$ and 
$D^{h/q}_{--}(z,\bpp)$ correspond to the transverse momentum dependent 
unpolarized fragmentation function $D_{h/q}(z,\pp)$, 
\be 
D^{h/q}_{++}(z,\bpp) = D^{h/q}_{--}(z,\bpp) = D_{h/q}(z,\pp) \,,
\label{unp-D}
\ee
whereas the non-diagonal elements 
\bea
D^{h/q}_{+-}(z,\bpp) &=& D^{h/q}_{+-}(z,\pp)\,e^{i\varphi} \,, \\
D^{h/q}_{-+}(z,\bpp) &=& D^{h/q}_{-+}(z,\pp)\,e^{-i\varphi}=
-D^{h/q}_{+-}(z,\pp)\,e^{-i\varphi}\;, \nonumber 
\label{pol-D} 
\eea
are related to the Collins fragmentation function 
$\Delta^N D_{h/q^\ua}(z,\pp)$ \cite{Anselmino:2005nn} by 

\be
\Delta^N D_{h/q^\ua}(z,\pp)=-2i\,D^{h/q}_{+-}(z,\pp)=2i\,D^{h/q}_{-+}(z,\pp)\,.
\label{Col-D}
\ee
The angle $\varphi$ in Eq.~(\ref{pol-D}) is the azimuthal 
angle identifying the direction of the 
observed hadron $h$ in the helicity frame of the fragmenting quark $q$.
Similar relations hold for the antiquark fragmentation functions, where one 
has to take into account a sign difference in $\varphi$ originating from the 
fact that 
the antiquark is chosen to move along the $-\hat z$ direction.
Finally, inserting Eqs.~(\ref{hel-amp})--(\ref{Col-D}) into 
Eq.~(\ref{general-belle}) and performing the sum over the quark helicities 
one obtains
\begin{widetext}
\bea
\frac{d\sigma ^{e^+e^-\to h_1 h_2 X}}
{dz_1\,dz_2\,d^2\bpo\,d^2\bpt\,d\cos\theta}&=&
 \frac{3\pi\alpha^2}{2s} \, \sum _q e_q^2 \, \Big\{
 (1+\cos^2\theta)\,D_{h_1/q}(z_1,\ppo)\,D_{h_2/\bar q}(z_2,\ppt)
\Big. \nonumber \\ & &
+ \Big. \frac{1}{4}\,\sin^2\theta\,\Delta ^N D _{h_1/q^\ua}(z_1,\ppo)\,
 \Delta ^N D _{h_2/\bar q^\ua}(z_2,\ppt)\,\cos(\varphi_1 + \varphi_2)\Big\}.
\label{belle}
\eea
\end{widetext}
Eq.~(\ref{belle}) shows that the study of the correlated production of two
hadrons (one for each jet) in unpolarized $e^+e^-$ collisions offers a direct
access to the Collins functions, both regarding their $z$ and $p_\perp$
dependences. So far, only data on the $z$ dependence are available.
Notice that by integrating over the intrinsic transverse momenta
$\bpo$ and $\bpt$ one recovers the usual unpolarized cross section,
\bea
\frac{d\sigma ^{e^+e^-\to h_1 h_2 X}}{dz_1\,dz_2\,d\cos\theta} 
&=&\\ &&\hspace*{-2cm}
 \frac{3\pi\alpha^2}{2s}\, (1+\cos^2\theta) \,
\sum _q e_q^2 \, D_{h_1/q}(z_1)\,D_{h_2/\bar q}(z_2)\;,\nonumber
\eea
having used
\be
\int d^2\bpp D_{h/q}(z,\pp) = D_{h/q}(z)\,.
\ee
To construct the physical observable measured by the Belle
Collaboration, we now perform a change of angular variables from
$(\varphi_1,\varphi_2)$ to  $(\varphi_1,\varphi_1+\varphi_2)$
and then integrate over the moduli of the intrinsic transverse momenta,
$\ppo$ and $\ppt$, and over the azimuthal angle $\varphi_1$. This leads to
\begin{widetext}
\bea
\frac{d\sigma ^{e^+e^-\to h_1 h_2 X}}
{dz_1\,dz_2\,d\cos\theta\,d(\varphi_1+\varphi_2)}&=&
 \frac{3\alpha^2}{4s} \, \sum _q e_q^2 \, \Big\{
 (1+\cos^2\theta)\,D_{h_1/q}(z_1)\,D_{h_2/\bar q}(z_2)
\frac{}{} \Big. \nonumber \\ & &
+\Big.\frac{1}{4}\,\sin^2\theta\,\cos(\varphi_1+\varphi_2)\,
\Delta ^N D _{h_1/q^\ua}(z_1)\,
\Delta ^N D _{h_2/\bar q^\ua}(z_2)\Big\}\,,
\label{int-Xs-belle}
\eea
\end{widetext}
where we have defined
\be
\int d^2\bpp \Delta ^N D_{h/q^\ua} (z,\pp) \equiv \Delta ^N D _{h/q^\ua}(z)\,.
\label{coll-mom}
\ee
By normalizing Eq.~(\ref{int-Xs-belle}) to the azimuthal averaged
cross section,
\bea
\langle d\sigma \rangle &\equiv& \frac{1}{2\pi} \,
\frac{d\sigma^{e^+e^-\to h_1 h_2 X}}{dz_1\,dz_2\,d\cos\theta} \nonumber\\
 && \hspace*{-1.6cm} =
\frac{3\alpha^2}{4s} \, \sum _q e_q^2 \,
 (1+\cos^2\theta)\,D_{h_1/q}(z_1)\,D_{h_2/\bar q}(z_2) \,, \label{phiav}
\eea
one has
\bea
A(z_1,z_2,\theta,\varphi_1 + \varphi_2) &\equiv& \frac{1}
{\langle d\sigma \rangle} \>
\frac{d\sigma ^{e^+e^-\to h_1 h_2 X}}{dz_1\,dz_2\,d\cos\theta\,
d(\varphi_1 + \varphi_2)} \nonumber\\
&& \hspace*{-1.5cm} =1+\frac{1}{4}\,\frac{\sin^2\theta}{1+\cos^2\theta}\,
\cos(\varphi_1+\varphi_2)\, \nonumber \\ && \hspace*{-1.5cm} \times
\frac{\sum_q e^2_q \, \Delta ^N D_{h_1/q^\ua}(z_1)\,
 \Delta ^N D_{h_2/\bar q^\ua}(z_2)}{\sum_q e^2_q D _{h_1/q}(z_1)\,
 D _{h_2/\bar q}(z_2)}\,\cdot \nonumber \\ && \; \label{A12g} 
\eea

Actually, Belle data are collected over a range of $\theta$ values, according
to the acceptance of the detector (see Eq. (\ref{belle-cuts})). Thus,
Eqs.~(\ref{int-Xs-belle}) and (\ref{phiav}) are integrated over the covered
$\theta$ range resulting in some specific $\langle \sin^2\theta \rangle$ and
$\langle 1+\cos^2\theta \rangle$ values.

Finally, to eliminate false asymmetries, the Belle Collaboration considers
the ratio of unlike-sign to like-sign pion pair production, $A_U$ and $A_L$,
given by
\bea
R &\equiv & \frac{A_U}{A_L} = \frac{1+\frac{1}{4}\,\cos(\varphi_1+\varphi_2)\,
  \frac{\langle \sin^2\theta \rangle}{\langle 1+\cos^2\theta \rangle}\,P_U}
       {1+\frac{1}{4}\,\cos(\varphi_1+\varphi_2)
  \frac{\langle \sin^2\theta \rangle}{\langle 1+\cos^2\theta \rangle}\,P_L}
\nonumber \\
& \simeq &  1+\frac{1}{4}\,\cos(\varphi_1+\varphi_2)  
  \frac{\langle \sin^2\theta \rangle}{\langle 1+\cos^2\theta \rangle} \,
  (P_U-P_L )\nonumber \\
&\equiv& 1+\cos(\varphi_1+\varphi_2)\,A_{12}(z_1,z_2)\,,
\label{R}
\eea
with
\begin{widetext}
\bea
&&
P_U= \frac{\sum_q e^2_q \;
    [\Delta ^N D _{\pi^+/q^\ua}(z_1)\,
     \Delta ^N D _{\pi^-/\bar q^\ua}(z_2) +
     \Delta ^N D _{\pi^-/q^\ua}(z_1)\,
     \Delta ^N D _{\pi^+/\bar q^\ua}(z_2)]}
{\sum_q e^2_q \;[D _{\pi^+/q}(z_1)\,D _{\pi^-/\bar q}(z_2) +
                 D _{\pi^-/q}(z_1)\,D _{\pi^+/\bar q}(z_2)]} \,,
\\
&&
P_L= \frac{\sum_q e^2_q \;
    [\Delta ^N D _{\pi^+/q^\ua}(z_1)\,
     \Delta ^N D _{\pi^+/\bar q^\ua}(z_2) +
     \Delta ^N D _{\pi^-/q^\ua}(z_1)\,
     \Delta ^N D _{\pi^-/\bar q^\ua}(z_2)]}
{\sum_q e^2_q \;[D _{\pi^+/q}(z_1)\,D _{\pi^+/\bar q}(z_2) +
                 D _{\pi^-/q}(z_1)\,D _{\pi^-/\bar q}(z_2)]}\,,
\\
&&
A_{12}(z_1,z_2)=\frac{1}{4}\frac{\langle \sin^2\theta \rangle}
{\langle 1+\cos^2\theta \rangle}\,(P_U-P_L )\,.
\label{A12}
\eea
For fitting purposes, it is convenient to re-express $P_U$ and $P_L$ in terms
of favoured and unfavoured fragmentation functions,
\bea
&& D_{\pi^+/u} = D_{\pi^+/\bar d} = D_{\pi^-/d} = D_{\pi^-/\bar u}
\equiv D_{\rm fav} \label{fav} \,,\\
&& D_{\pi^+/d} = D_{\pi^+/\bar u} = D_{\pi^-/u} = D_{\pi^-/\bar d}
= D_{\pi^\pm/s} =  D_{\pi^\pm/\bar s} \equiv D_{\rm unf}, \label{unf}
\eea
and similarly for the $\Delta^ND$, obtaining
\bea
&&P_U= \frac{\;
    [5\,\Delta ^N D _{\rm fav}(z_1)\,
        \Delta ^N D _{\rm fav}(z_2) +
     7\,\Delta ^N D _{\rm unf}(z_1)\,
        \Delta ^N D _{\rm unf}(z_2)]}
{   [5\,D _{\rm fav}(z_1)\,D _{\rm fav}(z_2) +
     7\,D _{\rm unf}(z_1)\,D _{\rm unf}(z_2)]} \,,
\label{PU}
\\
&&
P_L= \frac{
    [5\,\Delta ^N D _{\rm fav}(z_1)\,\Delta ^N D  _{\rm unf}(z_2) +
     5\,\Delta ^N D _{\rm unf}(z_1)\, \Delta ^N D  _{\rm fav}(z_2) +
     2\,\Delta ^N D _{\rm unf}(z_1)\,\Delta ^N D  _{\rm unf}(z_2) ]}
{   [5\,D _{\rm fav}(z_1)\,D _{\rm unf}(z_2) +
     5\,D _{\rm unf}(z_1)\,D _{\rm fav}(z_2) +
     2\,D _{\rm unf}(z_1)\,D _{\rm unf}(z_2)]}\,,\;\;\;\;
\label{PL}
\eea
\end{widetext}
having neglected heavy quark contributions. $P_U$ and $P_L$ are the same as
in Ref.~\cite{Efremov:2006qm}, remembering Eq.~(\ref{colmul}) and noticing
that
\bea
\label{Efre}
\Delta^ND_{h/q^\ua}(z)\!\!&=&\!\!\!\! \int \!\! d^2\bfp_\perp \Delta^N D_{h/q^\ua}(z,\pp)
\\
            \!\! &=&\!\!\!\! \int \!\!d^2\bfp_\perp \frac{2\pp}{z m_h} \; H_1^{\perp q}(z,\pp)
             = 4 \; H_1^{\perp(1/2)q}(z)\;.\nonumber 
\eea

\begin{figure}[t]
\begin{center}
\scalebox{0.30}{\input{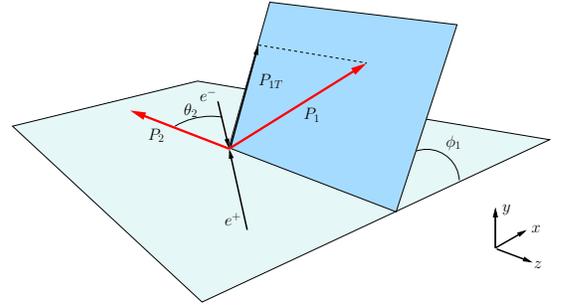}}
\caption{\small Three dimensional kinematics of the $e^+e^-\to h_1 h_2 \, X$
process. In this configuration the $\hat z$ direction is identified by the
momentum of the final hadron $h_2$, while $h_1$ is emitted at an azimuthal
angle $\phi_1$ with respect to the lepton-$h_2$ plane, defined as the
$\hat{xz}$ plane. }
\label{fig:epluseminus1}
\end{center}
\end{figure}

In addition, the Belle Collaboration presents a second set of data, analysed
in a different reference frame: following Ref.~\cite{Boer:1997mf}, one can
fix the $\hat z$-axis as given by the direction of the observed hadron $h_2$
and the $\hat{xz}$ plane as determined by the lepton and the $h_2$ directions.
There will then be another relevant plane, determined by $\hat z$ and the
direction of the other observed hadron $h_1$, at an angle $\phi_1$ with
respect to the $\hat{xz}$ plane. This kinematical configuration is shown in
Fig.~\ref{fig:epluseminus1}; it has the advantage that it does not require
the reconstruction of the quark direction.

However, in this case the kinematics is more complicated. At first order in
$p_\perp/(z\sqrt{s})$ one has
\bea
&&P_2=|P_2|(1,0,0,-1)\,, \\
&&q_2=\left( \frac{\sqrt{s}}{2},-\frac{\ppt}{z_2} \, \cos\varphi_2,
       -\frac{\ppt}{z_2} \, \sin\varphi_2,
       -\frac{\sqrt{s}}{2} \right) \,,\;\;\;\;\\
&&q_1=\left( \frac{\sqrt{s}}{2},\frac{\ppt}{z_2} \, \cos\varphi_2,
        \frac{\ppt}{z_2} \, \sin\varphi_2,
       \frac{\sqrt{s}}{2} \right)\,, \\
&&\bfP_1=\left( P_{1T}\cos\phi_1,P_{1T}\sin\phi_1,z_1\frac{\sqrt{s}}{2}
\right)\,, \\
&&\bfp_{\perp 1} = \left( P_{1T}\cos\phi_1-\frac{z_1}{z_2}\,\ppt\,
\cos\varphi_2\,,\right.\nonumber
\\ && \hspace*{1.2cm}\left.   P_{1T}\sin\phi_1-\frac{z_1}{z_2}\,\ppt\,\sin\varphi_2,\;
0 \right)\,.
\eea
Moreover, the elementary process $e^+e^- \to q \, \bar q$ does not occur in
general in the $\hat{xz}$ plane, and thus the helicity scattering amplitudes
involve an azimuthal phase $\varphi_2$.
One can still perform an exact calculation, using the general approach
discussed in Ref.~\cite{Anselmino:2005sh}. A detailed description will be
presented in a forthcoming paper~\cite{sidis-general}. We give here only the
results valid at ${\cal O}(\pp/z\sqrt s)$.
The analogue of Eq.~(\ref{belle}) now reads
\begin{widetext}
\bea
\frac{d\sigma ^{e^+e^-\to h_1 h_2 X}}
{dz_1\,dz_2\,d^2\bpo\,d^2\bpt\,d\cos\theta_2}&=&
 \frac{3\pi\alpha^2}{2s} \, \sum _q e_q^2 \, \Big\{
 (1+\cos^2\theta_2)\,D_{h_1/q}(z_1,\ppo)\,D_{h_2/\bar q}(z_2,\ppt)
\frac{}{} \Big.  \!\!\! \label{belle2} \\ & &
+ \Big. \frac{1}{4}\,\sin^2\theta_2\,\Delta ^N D _{h_1/q^\ua}(z_1,\ppo)\,
 \Delta ^N D _{h_2/\bar q^\ua}(z_2,\ppt)\,
\cos(2\varphi_2 + \phi_{q}^{h{_1}})\Big\}\,,\nonumber
\eea
\end{widetext}
where $\phi_{q}^{h_1}$ is the azimuthal angle of the detected hadron $h_1$
around the direction of the parent fragmenting quark, $q$. Technically,
$\phi_{q}^{h_1}$ is the azimuthal angle of $\bpo$ in the helicity frame of
$q$. It can be expressed in terms of the integration variables we are using,
$\bpt$ and $P_{1T}$. At lowest order in $\pp/(z\sqrt{s})$ we have
\bea
&&\cos\phi_{q}^{h_1} = \frac{P_{1T}}{\ppo} \,
\cos(\phi_1-\varphi_2) - \frac{z_1}{z_2} \, \frac{\ppt}{\ppo} \,,\\
&&\sin\phi_{q}^{h_1}=
\frac{P_{1T}}{\ppo} \, \sin(\phi_1-\varphi_2) \;.
\eea
Integrating Eq.~(\ref{belle2}) over $\bpt$ and $P_{1T}$, but not over $\phi_1$,
and normalizing to the azimuthal averaged unpolarized cross section
(\ref{phiav}), we obtain the analogue of Eq.~(\ref{A12g}),
\bea
A(z_1,z_2,\theta_2,\phi_1) &=& 1+\frac{1}{\pi}\,\frac{z_1\,z_2}{z_1^2+z_2^2}\,
\frac{\sin^2\theta_2}{1+\cos^2\theta_2}\,\cos(2\,\phi_1)\,
\nonumber \\ &\times& 
\frac{\sum_q e^2_q \, \Delta ^N D  _{h_1/q^\ua}(z_1)\,
 \Delta ^N D  _{h_2/\bar q^\ua}(z_2)}{\sum_q e^2_q D _{h_1/q}(z_1)\,
 D _{h_2/\bar q}(z_2)}\,,\nonumber\\&&
\eea
in agreement with Ref.~\cite{Efremov:2006qm} taking into account the different
notations, Eqs.~(\ref{colmul}) and (\ref{Efre}).
%
%
%%%%%%%%%%%%%%%%%%%%%%%%%%%%%%%%%%%%%%%%%%%%%%%%%%%%%%%%%%%%%%%%%%%%%%%%%%%
%%%%%%%%%%%%%%%%%%%%%%%%%%%%%%%%%%%%%%%%%%%%%%%%%%%%%%%%%%%%%%%%%%%%%%%%%%%
%
%

Finally, Eq.~(\ref{R}) becomes in this configuration
\be
R \simeq 1+\cos(2\,\phi_1)A_0(z_1,z_2)\;, \label{R0}
\ee
with
\be
A_0(z_1,z_2)=\frac{1}{\pi} \, \frac{z_1\,z_2}{z_1^2+z_2^2}
\, \frac{\langle \sin^2\theta_2 \rangle}
{\langle 1+\cos^2\theta_2 \rangle}\,(P_U-P_L )\,,
\label{A0}
\ee
where $P_U$ and $P_L$ are the same as defined in Eqs.~(\ref{PU}) and
(\ref{PL}).

\section{\label{fit} Transversity and Collins functions from a global fit}

We can now pursue our strategy of gathering simultaneous information
on the transversity distribution function $\Delta_Tq(x,\kt)$ and
the Collins fragmentation function $\Delta ^N D_{h/\qup}(z,\pp)$. To such a
purpose we perform a global best fit analysis of experimental data involving
these functions, namely the data from the SIDIS measurements by
the HERMES~\cite{Airapetian:2004tw,HERMES:proceedings} and COMPASS~\cite{Ageev:2006da}
Collaborations, and the data from unpolarized $e^+e^-\to h_1h_2\,X$  processes
by the Belle Collaboration~\cite{Abe:2005zx}.
%
%%%%%%%%%%%%%%%%%%%%%%%%%%%%%%%%%%%%%%%%%%%%%%%%%%%%%%%%%%%%%%%%%%%%%%%%%%%%%%%
\begin{table}[t]
\caption{Best values of the free parameters for the $u$ and $d$
transversity distribution functions and for the favored and unfavored
Collins
fragmentation functions, Eqs.~(\ref{tr-funct})-(\ref{h-funct}), as obtained by 
simultaneously fitting HERMES and COMPASS data on the 
$A_{UT}^{\sin(\phi_S+\phi_h)}$ asymmetry and the Belle data on the $A_{12}$ 
asymmetry, Eq.~(\ref{A12}), proportional to $\cos(\varphi_1+\varphi_2)$. 
Notice that
the errors generated by MINUIT are strongly correlated, and should not
be taken
at face value. The significant fluctuations in our results are shown by the
shaded areas in Figs.~\ref{fig:hermes}, \ref{fig:compass} and
\ref{fig:belle},
as explained in the text. 
The values of $\avk \T = \avk $ and
$\langle \pp^2 \rangle$ are fixed, according to Eq.~(\ref{ktpp}).
\label{fitpar12}}
\begin{center}
\begin{tabular}{lllllll}
\hline
\hline
FIT I ($A_{12}$) & ~& ~& $\chi ^2/{\rm d.o.f.}\;\;$ = & $0.81$ & ~ & ~\\
%~&~&~&~&~&~&~\\
\hline
Transversity &
$N_{u}^T$ &=&  $0.48  \pm  0.09$ ~~~& $N_{d}^T$ &=&  $\!-0.62  \pm
0.30$ \\
distribution &
$\alpha$ &=&  $1.14  \pm  0.68$ & $\beta$  &=&  $4.74  \pm  5.45$   \\
function &
$\avk\T\!$ &=&  0.25 GeV$^2$ &~&~&~\\
%~&~&~&~&~&~&~\\
%\hline
~&~&~&~&~&~&~\\
Collins &
$N_{\rm fav}^C$  &=& $0.35  \pm  0.16$ & $N_{\rm unf}^C$   &=& $\!-0.85 
\pm
0.36$ \\
fragmentation~~ &
$\gamma$  &=& $1.14  \pm  0.38$ & $\delta$   &=& $0.14 \pm  0.36$    \\
function &
$\avp$  &=& 0.20 GeV$^2$ & $M^2$  &=&  $0.70 \pm 0.65$\\
~&~&~&~&~&~&~~~~~~~~~GeV$^2$\\
%\hline
%~&~&~&~&~&~&~\\
%~ & ~& ~& $\chi ^2/{\rm d.o.f.}\;\;$ = & $0.81$ & ~ & ~\\
%~&~&~&~&~&~&~\\
\hline
\hline
\end{tabular}
\end{center}
\end{table}
%%%%%%%%%%%%%%%%%%%%%%%%%%%%%%%%%%%%%%%%%%%%%%%%%%%%%%%%%%%%%%%%%%%%%%%%%%%%%%%
%

$\Delta_Tq(x,\kt)$ and $\Delta ^N D_{h/\qup}(z,\pp)$ are parameterized
as shown in Eqs.~(\ref{tr-funct})--(\ref{h-funct}). Considering the
scarcity of data, in order to minimize the number of parameters, we
assume flavor independent values of $\alpha$ and $\beta$ and, similarly,
we assume that $\gamma$ and $\delta$ are the same for favored and
unfavored Collins fragmentation functions, Eqs.~(\ref{fav}) and
(\ref{unf}); we then remain with a total number of 9 parameters. Their
values, as determined through our global best fit are shown in Table
\ref{fitpar12} and \ref{fitpar0}, together with the errors estimated by
MINUIT.

\begin{figure}[t]
\vspace*{-1cm}
\hspace*{1.4cm}
\includegraphics[width=0.42\textwidth,height=0.45\textwidth,bb= 10 140 540 660,
angle=-90]
{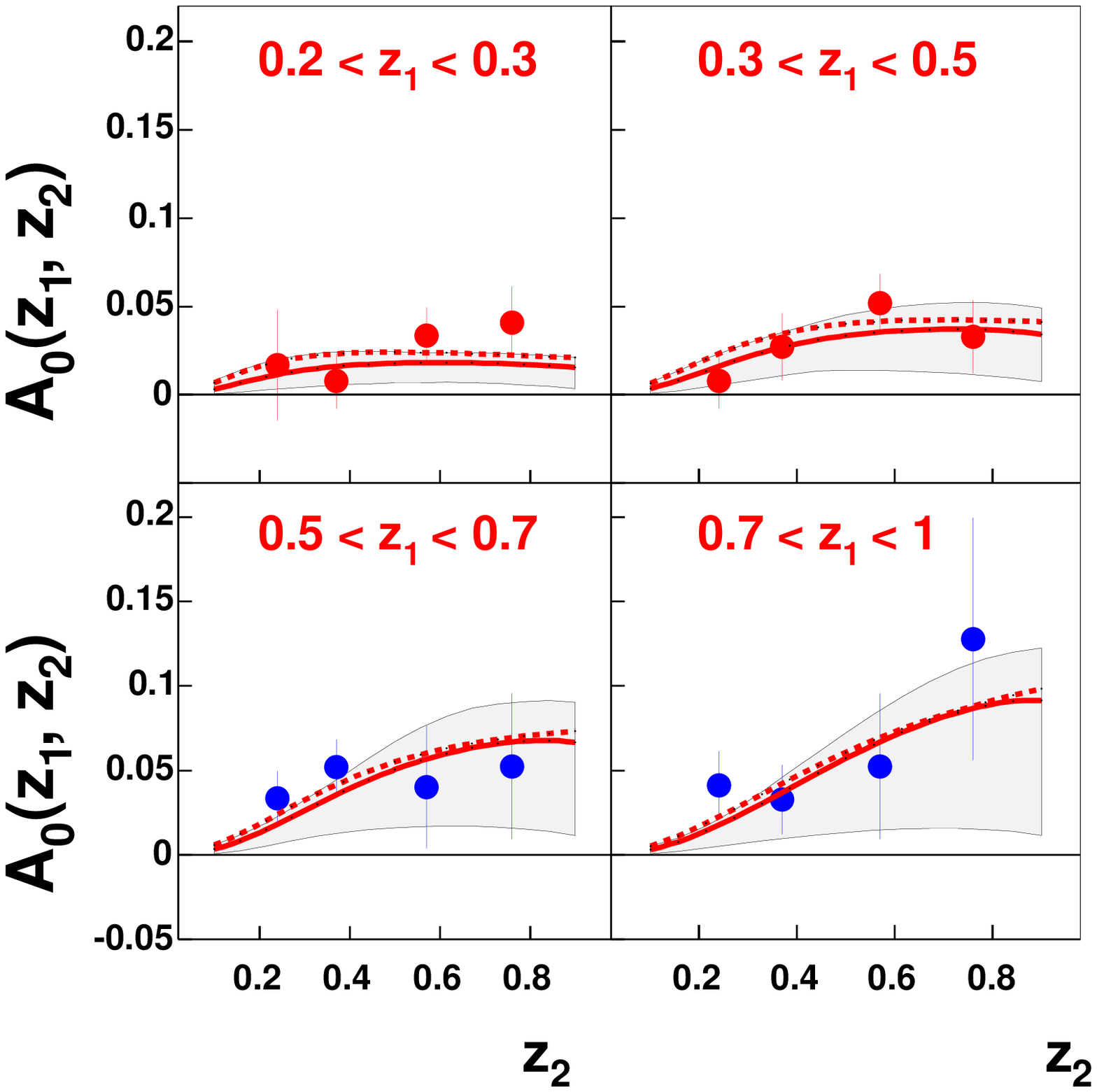} 
\\ \hspace*{1.4cm}
\includegraphics[width=0.42\textwidth,height=0.45\textwidth,bb= 10 140 540 660,
angle=-90]
{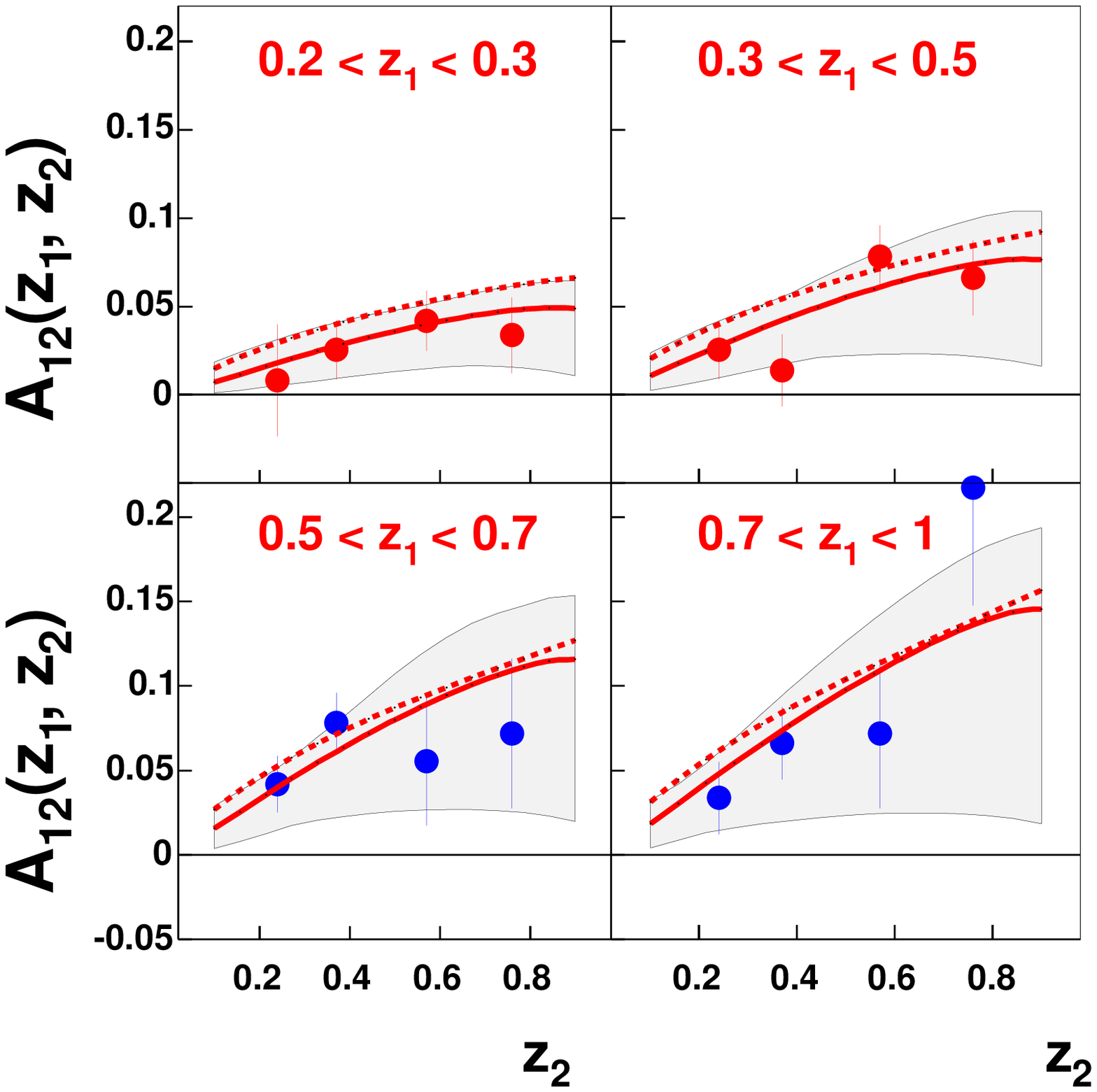}
\caption{\label{fig:belle}
The experimental data on two different azimuthal correlations in unpolarized
$e^+e^- \to h_1 h_2 \, X$ processes, as measured by Belle Collaboration
\cite{Abe:2005zx}, are compared to the curves obtained from Eqs.~(\ref{A12})
[$A_{12}$] and (\ref{A0}) [$A_0$] with the parameterizations of 
Eqs.~(\ref{coll-funct}),
(\ref{NC}) and (\ref{h-funct}).
The solid lines correspond to the parameters given in 
Table~\ref{fitpar12}, obtained by fitting the $A_{12}$ asymmetry; 
the shaded area corresponds to the theoretical uncertainty on these 
parameters, as explained in the text.
The dashed lines correspond to the  parameters given in  Table~\ref{fitpar0} 
obtained by fitting the $A_{0}$ asymmetry. The agreement between the results 
obtained from the two fits shows the consistency between the two sets of Belle 
data and the solidity of our analysis. }
\end{figure}

\begin{figure}[H]
\vspace*{-0.9cm}
\hspace{0.8cm}
\includegraphics[width=0.38\textwidth,,bb= 10 140 540 660,angle=-90]
{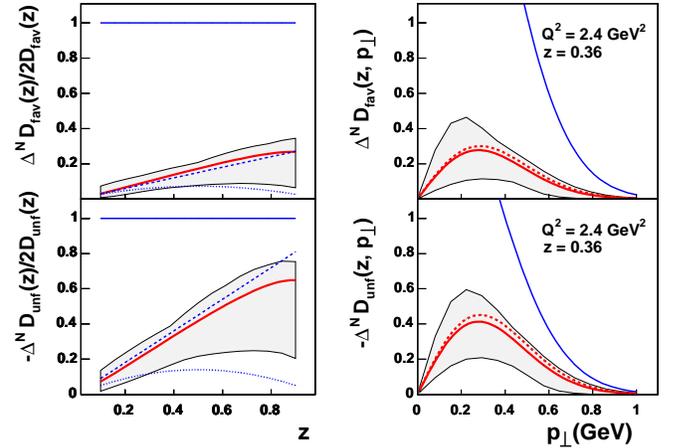}
\caption{\label{fig:coll}
Favored and unfavored Collins fragmentation functions as determined through
our global best fit. In the left panel we show the $z$ dependence of the
$\pp$ integrated Collins functions defined in Eq.~(\ref{coll-mom}) and
normalized to twice the corresponding unpolarized fragmentation functions; 
we compare them to the results of Refs.~\cite{Efremov:2006qm} (dashed line) 
and \cite{Vogelsang:2005cs} (dotted line). In the right panel we show the
$\pp$ dependence of the Collins functions defined in Eq.~(\ref{coll-funct}),
at a fixed value of $z$. The $Q^2$ value is 2.4 GeV$^2$, having assumed 
that the $Q^2$ evolution of $\Delta^N D$ is the same as that of $D$.
The solid lines show the results based on the 
parameters of Table \ref{fitpar12}, while the dashed ones show the results
corresponding to the parameters of Table \ref{fitpar0}. In all cases we also 
show the positivity bound (\ref{bound}) (upper lines).}
\end{figure}

As the two different sets of Belle data are based on a different
analysis of the same experimental events, they are strongly correlated.
Therefore, we have treated them separately in our combined analysis of
the HERMES, COMPASS and Belle data; the best fit values of Table
\ref{fitpar12} are obtained by fitting the SIDIS results together with
the Belle data on the $\cos(\varphi_1 + \varphi_2)$ dependence,
Eq. (\ref{R}), while the values in Table \ref{fitpar0} originate from
the Belle data on the $\cos(2\phi_1)$ dependence, Eq.~(\ref{R0}). 
We notice that the two sets of resulting best fit parameters are in full
agreement within the uncertainties; this gives a good check of the
consistency of the measurements and the stability of our analysis. In
the sequel we shall present results and predictions based on the
values of Table~\ref{fitpar12}; the corresponding results based on the
values of Table~\ref{fitpar0}  are hardly distinguishable (examples of this 
are shown explicitely in Fig.~\ref{fig:belle} and in Fig.~\ref{fig:coll}, 
right panel).

\begin{figure}[t]
\vspace*{-0.8cm}
\includegraphics[width=0.36\textwidth,bb= 10 140 540 660,angle=-90]
{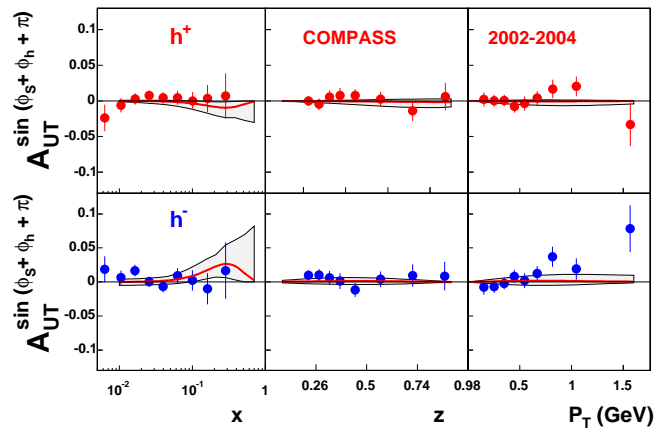} \hskip 1.85cm
\caption{\label{fig:compass}
The measurements of $A_{_{UT}}^{\sin(\phi_S+\phi_h)}$, for the production of
positively and negatively charged hadrons, from the COMPASS experiment
operating on a deuterium target~\cite{Ageev:2006da} are compared to the
curves obtained from Eq.~(\ref{sin-asym-final}) with the parameterizations
of Eqs.~(\ref{tr-funct})-(\ref{h-funct}), and the parameter values,
determined through our global best fit, given in Table~\ref{fitpar12}.
The shaded area corresponds to the theoretical uncertainty on the parameters,
as explained in the text. Notice the extra $\pi$ phase in addition to
$\phi_S + \phi_h$ in the figure label, to keep into account the different
choice of the Collins angle, with respect to Trento \cite{Bacchetta:2004jz}
and HERMES conventions, adopted by COMPASS Collaboration.}
\end{figure}

Our best fits of the experimental data from HERMES, COMPASS and Belle
are shown in Figs.~\ref{fig:hermes}, \ref{fig:compass} and
\ref{fig:belle} respectively. The central curves correspond to the central 
values of the parameters in Table~\ref{fitpar12}, while the shaded areas 
correspond to one-sigma deviation at 90\% CL and are calculated using the 
errors and the parameter correlation matrix generated by MINUIT, minimizing 
and maximizing the function under consideration, in a 9-dimensional
parameter space hyper-volume corresponding to one-sigma deviation.

\begin{figure}[H]
\vspace*{-0.9cm}
\hspace{0.8cm}
\includegraphics[width=0.38\textwidth,bb= 10 140 540 660,angle=-90]
{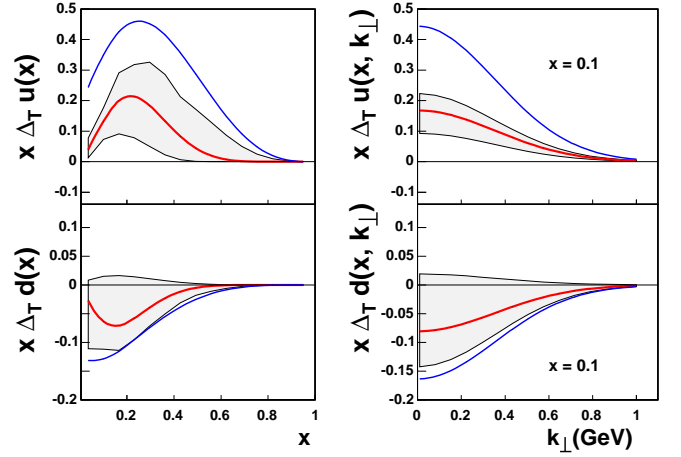}
\caption{\label{fig:transv}
The transversity distribution functions for $u$ and $d$ quarks as determined
through our global best fit. In the left panel, $x\,\Delta _T u(x)$ (upper
plot) and $x\,\Delta _T d(x)$ (lower plot), see Eq.~(\ref{h1}), are shown as
functions of $x$ and $Q^2 = 2.4$ GeV$^2$. The Soffer bound 
\cite{Soffer:1994ww} is also shown for comparison (bold blue line).
In the right panel we present the unintegrated transversity
distributions, $x\,\Delta _T u(x,\kt)$ (upper plot) and $x\,\Delta _T d(x,\kt)$
(lower plot), as defined in Eq.~(\ref{tr-funct}), as functions of $\kt$ at a
fixed value of $x$. Notice that this $\kt$ dependence is not obtained from the
fit, but it has been chosen to be the same as that of the unpolarized
distribution functions: we plot it in order to show its uncertainty (shaded
area), due to the uncertainty in the determination of the free parameters.
}
\end{figure}

The transversity distribution functions $\Delta_T u(x,\kt)$ and
$\Delta_T d(x,\kt)$ as resulting from our best fit --
Eqs.~(\ref{tr-funct})--(\ref{h-funct}) and Table \ref{fitpar12} -- are plotted
as a function of $x$ and $\kt$ in Fig.~\ref{fig:transv}; for comparison,
the Soffer bound of Eq.~(\ref{soffer}) is also shown, as a bold line.
The solid central line corresponds to the central values in 
Table~\ref{fitpar12} and the shaded area corresponds to the uncertainty in 
the parameter values, as explained above.

Similarly, the resulting Collins functions $\Delta^ND_{\rm fav}(z,\pp)$
and $\Delta^ND_{\rm unf}(z,\pp)$ are plotted as a function of $z$ -- 
integrated over $d^2\bfp_\perp$, Eq.~(\ref{Efre}), and  
normalized to twice the unpolarized fragmentation functions --
and as a function of $\pp$ in Fig.~\ref{fig:coll}; for comparison, we also
show the Collins functions from Refs.~\cite{Efremov:2006qm} and
\cite{Vogelsang:2005cs}, respectively as dashed and dotted lines 
(left panels), 
and the corresponding positivity bound (\ref{bound}).
The dashed lines in the right panels show the results
corresponding to the parameters of Table \ref{fitpar0}.
\\
%
%%%%%%%%%%%%%%%%%%%%%%%%%%%%%%%%%%%%%%%%%%%%%%%%%%%%%%%%%%%%%%%%%%%%%%%%%%%%%%%
\begin{table}[t]
\caption{Best values of the free parameters for the $u$ and $d$
transversity distribution functions and for the favored and unfavored
Collins fragmentation functions, Eqs.~(\ref{tr-funct})-(\ref{h-funct}), 
as obtained by simultaneously fitting HERMES and COMPASS data on the 
$A_{UT}^{\sin(\phi_S+\phi_h)}$ asymmetry and the Belle data on the $A_{0}$ 
asymmetry, Eq.~(\ref{A0}), proportional to $\cos(2\phi_1)$. 
Notice that
the errors generated by MINUIT are strongly correlated, and should not
be taken
at face value. The significant fluctuations in our results are shown by the
shaded areas in Figs.~\ref{fig:hermes}, \ref{fig:compass} and
\ref{fig:belle},
as explained in the text. 
The values of $\avk \T = \avk $ and
$\langle \pp^2 \rangle$ are fixed, according to Eq.~(\ref{ktpp}).
\label{fitpar0}}
\begin{center}
\begin{tabular}{lllllll}
\hline
\hline
FIT II ($A_{0}$) & ~& ~& $\chi ^2/{\rm d.o.f.}\;\;$ = & $0.77$ & ~ & ~\\
%~&~&~&~&~&~&~\\
\hline
Transversity &
$N_{u}^T$ &=&  $0.42  \pm  0.09$ ~~~& $N_{d}^T$ &=&  $\!-0.53  \pm
0.28$ \\
distribution &
$\alpha$ &=&  $1.20  \pm  0.83$ & $\beta$  &=&  $5.09  \pm  5.87$   \\
function &
$\avk\T\!$ &=&  0.25 GeV$^2$ & & & \\
%~&~&~&~&~&~&~\\
%\hline
~&~&~&~&~&~&~\\
Collins &
$N_{\rm fav}^C$  &=& $0.41  \pm  0.10$ & $N_{\rm unf}^C$   &=& $\!-0.99 
\pm
1.24$ \\
fragmentation~ &
$\gamma$  &=& $0.81  \pm  0.40$ & $\delta$   &=& $0.02 \pm  0.37$    \\
function &
$\avp$  &=& 0.20 GeV$^2$ & $M^2$  &=&  $0.88 \pm 1.15$\\
~&~&~&~&~&~&~~~~~~~~~GeV$^2$\\
\hline
\hline
\end{tabular}
\end{center}
\end{table}
%%%%%%%%%%%%%%%%%%%%%%%%%%%%%%%%%%%%%%%%%%%%%%%%%%%%%%%%%%%%%%%%%%%%%%%%%%%%%%%
%
A few comments are in order.
\begin{itemize}
\item 
In Fig. \ref{fig:transv} we show the extracted transversity distribution
for $u$ and $d$ quarks. The $x$ dependence is based on the simple
parameterization assumed in Eqs. (\ref{tr-funct}) and (\ref{NT}),
which contain $N^{\T}_q, \alpha$ and $\beta$ as free parameters; our result
represents the first extraction ever of the transversity distributions
$\Delta_T u(x)$ and $\Delta_T d(x)$.
\item
The $\kt$ dependence has been assumed to be the same as for the unpolarized
distributions. The flavor dependence is contained in the coefficients
$N^{\T}_q$ and in the proportionality of $\Delta_Tq(x)$ to
$[q(x) + \Delta q(x)]/2 = q_+^+(x)$, the number density of quarks with
positive helicity inside a positive helicity proton.
\item
Our results show that the transversity distribution is positive for $u$ quarks
and negative for $d$ quarks; the magnitude of $\Delta_Tu$ is larger than that
of $\Delta_Td$, while they are both significantly smaller than the
corresponding Soffer bound.
%
%%%%%%%%%%%%%%%%%%%%%%%%%%%%%%%%%%%%%%%%%%%%%%%%%%%%%%%%%%%%%%%%%%%%%%%%%%%%%%%
\begin{figure}[t]
\vspace*{-0.8cm}
\includegraphics[width=0.36\textwidth,bb= 10 140 540 660,angle=-90]
{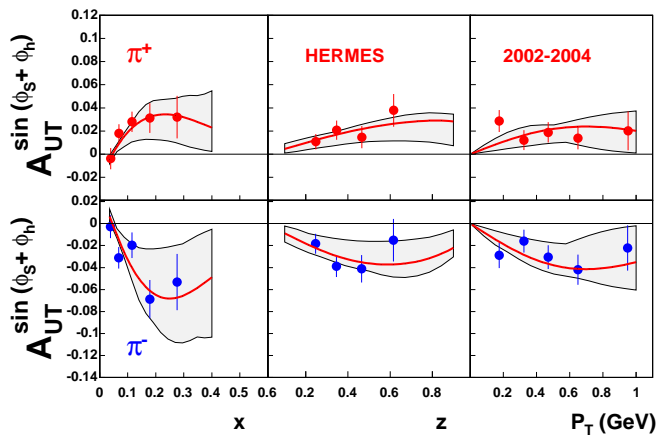} \hskip 2.85cm
\caption{\label{fig:hermes}
HERMES experimental data \cite{Airapetian:2004tw,HERMES:proceedings} on the 
azimuthal
asymmetry $A_{_{UT}}^{\sin(\phi_S+\phi_h)}$ for $\pi^\pm$ production
are compared to the curves obtained from Eq.~(\ref{sin-asym-final}) with
the parameterizations of Eqs.~(\ref{tr-funct})-(\ref{h-funct}), and the
parameter values, determined through our global best fit, given in
Table~\ref{fitpar12}. The shaded area corresponds to the theoretical
uncertainty on the parameters, as explained in the text.}
\end{figure}
%%%%%%%%%%%%%%%%%%%%%%%%%%%%%%%%%%%%%%%%%%%%%%%%%%%%%%%%%%%%%%%%%%%%%%%%%%%%%%%
%
\item
The shaded regions in Fig.~\ref{fig:transv} show that both $\Delta_T
u(x,\kt)$ and $\Delta_T d(x,\kt)$ are, considering the limited amount of
data, already well determined. It is worth noticing that while
the HERMES data alone tightly constrain the transversity distribution of
$u$ quarks, the addition of COMPASS data to the fit allows to better
constrain the transversity distribution function of $d$ quarks.
We have checked that fitting only HERMES and Belle data, ignoring the
COMPASS results, still leads to a similar good $\chi^2/{\rm d.o.f.}$; the
resulting functions would give a slightly worse description -- when
compared to the global fit -- of the $x$ dependence of
$A_{_{UT}}^{\sin(\phi_S+\phi_h)}$, as measured by COMPASS. This is
mainly related to a less stringent determination of $\Delta_T d(x,\kt)$
in absence of deuteron target data. Although their measured azimuthal
asymmetry is very small, the inclusion of COMPASS data significantly
contributes to the extraction of the transversity distributions.
Different fitting procedures were earlier attempted, for
example by fixing $\Delta_Tq = \Delta q$ or $\Delta_Tq = q_+^+$
\cite{Prokudin:2006fa}: they lead to a slightly worse description of Belle
data.
\item
The extracted Collins functions are shown in Fig. \ref{fig:coll}; they agree
with similar extractions previously obtained in the literature
\cite{Efremov:2006qm,Vogelsang:2005cs}. The shaded areas indicate well
constrained Collins functions for $u$ and $d$ quarks in the large (valence)
$z$ region, much smaller than their corresponding positivity bound.
\item
We note once more that, in analyzing SIDIS data, we have neglected the
contributions of the sea quarks and antiquarks (assuming the corresponding
transversity distributions in a proton to vanish), taking into account only
$u$ and $d$ flavors. In analyzing Belle data and introducing the favored and
unfavored Collins fragmentation functions, we have considered the contributions
of $u, d$ and $s$ quarks, all abundantly produced in the $e^+e^-$ annihilation
at $\sqrt s \simeq$ 10 GeV.
\item
The partonic distribution and fragmentation functions are taken from
Refs.~\cite{Gluck:1998xa,Gluck:2000dy} and \cite{Kretzer:2000yf}.
The QCD evolution is taken into account in the unpolarized distributions,
in the unpolarized fragmentation functions and, following
Ref.~\cite{Martin:1997rz}, for the transversity distributions.
\end{itemize}

%%%%%%%%%%%%%%%%%%%%%%%%%%%%%%%%%%%%%%%%%%%%%%%%%%%%%%%%%%%%%%%%%%%%%%%%%%%%%%%
%%%%%%%%%%%%%%%%%%%%%%%%%%%%%%%%%%%%%%%%%%%%%%%%%%%%%%%%%%%%%%%%%%%%%%%%%%%%%%%
%
%%%%%%%%%%%%%%%%%%%%%%%%%%%%%%%%%%%%%%%%%%%%%%%%%%%%%%%%%%%%%%%%%%%%%%%%%%%%%%%
%%%%%%%%%%%%%%%%%%%%%%%%%%%%%%%%%%%%%%%%%%%%%%%%%%%%%%%%%%%%%%%%%%%%%%%%%%%%%%%
%
%
%%%%%%%%%%%%%%%%%%%%%%%%%%%%%%%%%%%%%%%%%%%%%%%%%%%%%%%%%%%%%%%%%%%%%%%%%%%%%%%
%%%%%%%%%%%%%%%%%%%%%%%%%%%%%%%%%%%%%%%%%%%%%%%%%%%%%%%%%%%%%%%%%%%%%%%%%%%%%%%
%
%%%%%%%%%%%%%%%%%%%%%%%%%%%%%%%%%%%%%%%%%%%%%%%%%%%%%%%%%%%%%%%%%%%%%%%%%%%%%%%
%%%%%%%%%%%%%%%%%%%%%%%%%%%%%%%%%%%%%%%%%%%%%%%%%%%%%%%%%%%%%%%%%%%%%%%%%%%%%%%
%

Finally, we explicitly list, for clarity and completeness, the kinematical
cuts we have imposed in numerical integrations, according to the setup
of the HERMES experiment:
\bea
&& 0.2 \le z_h \le 0.7 \,, \quad\quad  0.023 \le \xb \le 0.4\,, \nonumber \\
&& 0.1 \le y \le 0.85 \,,  \quad\quad  Q^2 \ge 1 \;{\rm GeV}^2 \,, \\
&& W^2 \ge 10\; {\rm GeV}^2\,, \quad\quad 2 \le E_h \le 15 \; {\rm GeV}\;,
\nonumber
\label{hermes-cuts}
\eea
the COMPASS experiment:
\bea
&&0.2 \le z_h \le 1 \,,\quad\quad
  0.1 \le y \le 0.9 \,, \nonumber \\
&&Q^2 \ge 1 \;{\rm GeV}^2 \,,\quad\quad
  W^2 \ge 25\; {\rm GeV}^2 \;,
\label{compass-cuts}
\eea
and the Belle experiment
\be
-0.6 \le \cos \theta_{\rm lab} \le 0.9 \,,
\quad\quad 
Q_T \le 3.5 \;{\rm GeV} \;,
\label{belle-cuts}
\ee
where $\theta_{\rm lab}$ is the polar production angle in the laboratory
frame (related to the scattering angles $\theta$ and $\theta_2$ used in this
paper) and $Q_T$ is the transverse momentum of the virtual photon from the
$e^+e^-$ annihilation in the rest frame of the hadron pair \cite{Boer:1997mf}.

\section{\label{pred} Predictions for ongoing and future experiments}

We can now use the transversity distributions and the Collins functions
we have obtained from fitting the available HERMES, COMPASS and Belle
data, see Table~\ref{fitpar12}, 
to give predictions for new measurements planned by COMPASS and JLab
Collaborations.

\begin{figure}[ht]
\vspace*{-0.8cm}
\includegraphics[width=0.35\textwidth,bb= 10 140 540 660,angle=-90]
{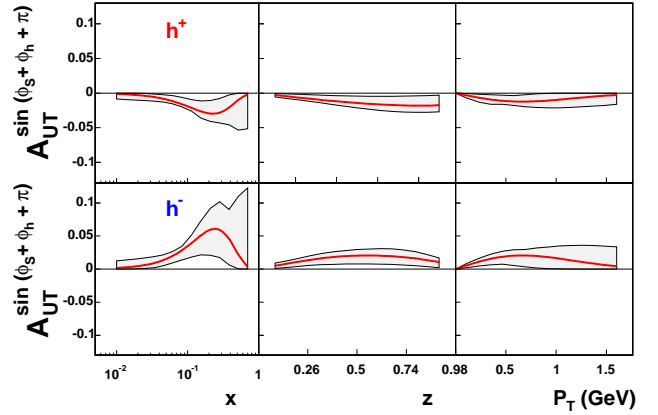}
\caption{\label{fig:compass-proton}
Predictions for the single spin asymmetry $A_{_{UT}}^{\sin(\phi_S+\phi_h)}$
as it will be measured by the COMPASS experiment operating with a
transversely polarized hydrogen target. For the extra $\pi$ phase in the 
figure label see the caption of Fig. 5.}
\end{figure}

The transverse single spin asymmetry $A_{_{UT}}^{\sin(\phi_S+\phi_h)}$
will be measured by the COMPASS experiment operating with a polarized hydrogen
target (rather than a deuterium one). In Fig.~\ref{fig:compass-proton}
we show our predictions, obtained by adopting the same experimental cuts
which were used for the deuterium target, see Eq.~(\ref{compass-cuts}).
Notice that this asymmetry is found to be sizeable, up to 5\% in size.
%%%%%%%%%%%%%%%%%%%%%%%%%%%%%%%%%%%%%%%%%%%%%%%%%%%%%%%%%%%%%%%%%%%%%%%%%%%%%%%
%%%%%%%%%%%%%%%%%%%%%%%%%%%%%%%%%%%%%%%%%%%%%%%%%%%%%%%%%%%%%%%%%%%%%%%%%%%%%%%
%
%%%%%%%%%%%%%%%%%%%%%%%%%%%%%%%%%%%%%%%%%%%%%%%%%%%%%%%%%%%%%%%%%%%%%%%%%%%%%%%
%
%%%%%%%%%%%%%%%%%%%%%%%%%%%%%%%%%%%%%%%%%%%%%%%%%%%%%%%%%%%%%%%%%%%%%%%%%%%%%%%
%
%%%%%%%%%%%%%%%%%%%%%%%%%%%%%%%%%%%%%%%%%%%%%%%%%%%%%%%%%%%%%%%%%%%%%%%%%%%%%%%
\begin{figure}[t]
\vspace*{-0.7cm}
\includegraphics[width=0.33\textwidth,bb= 10 140 540 660,angle=-90]
{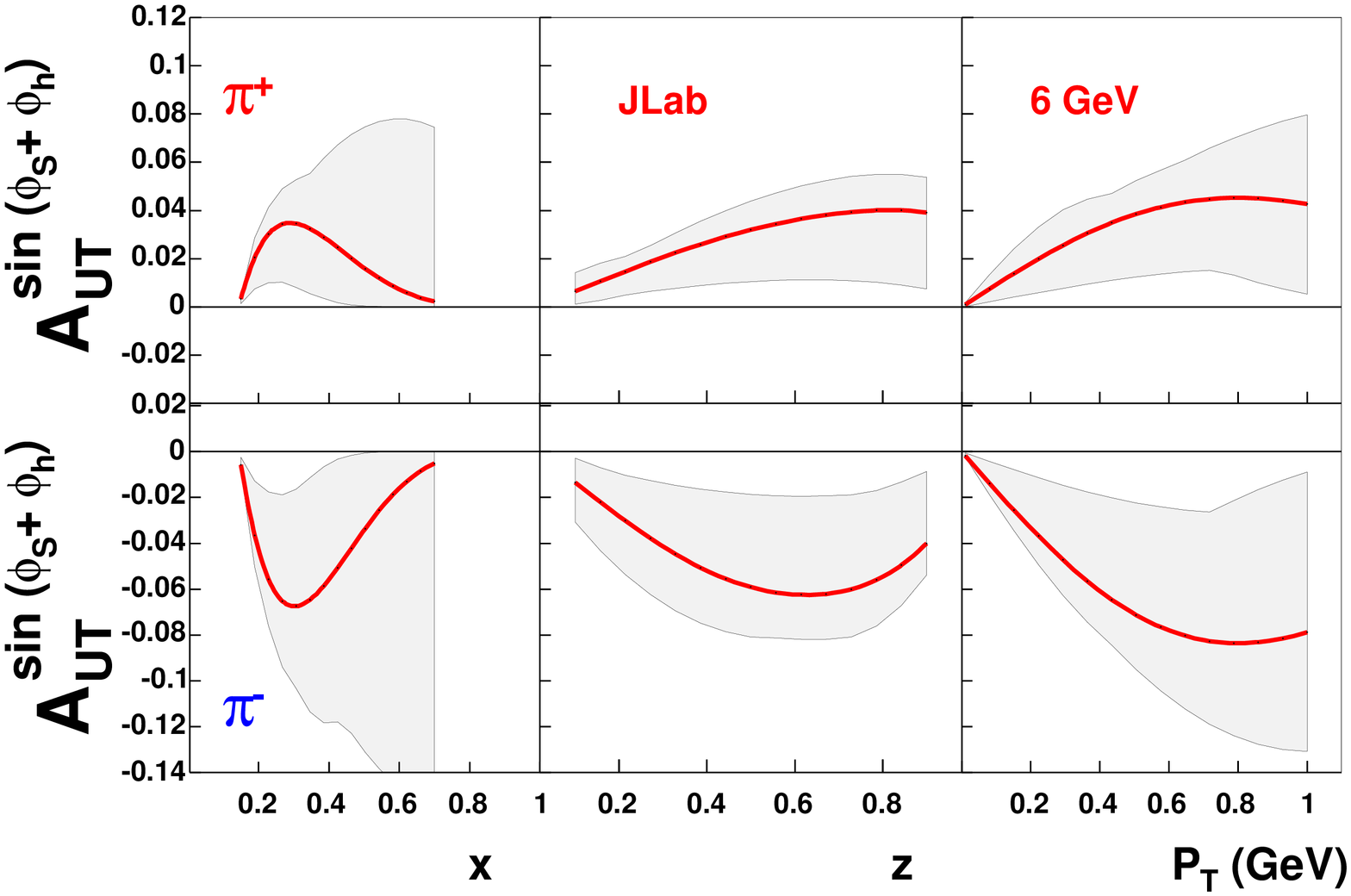} \\
\includegraphics[width=0.33\textwidth,bb= 10 140 540 660,angle=-90]
{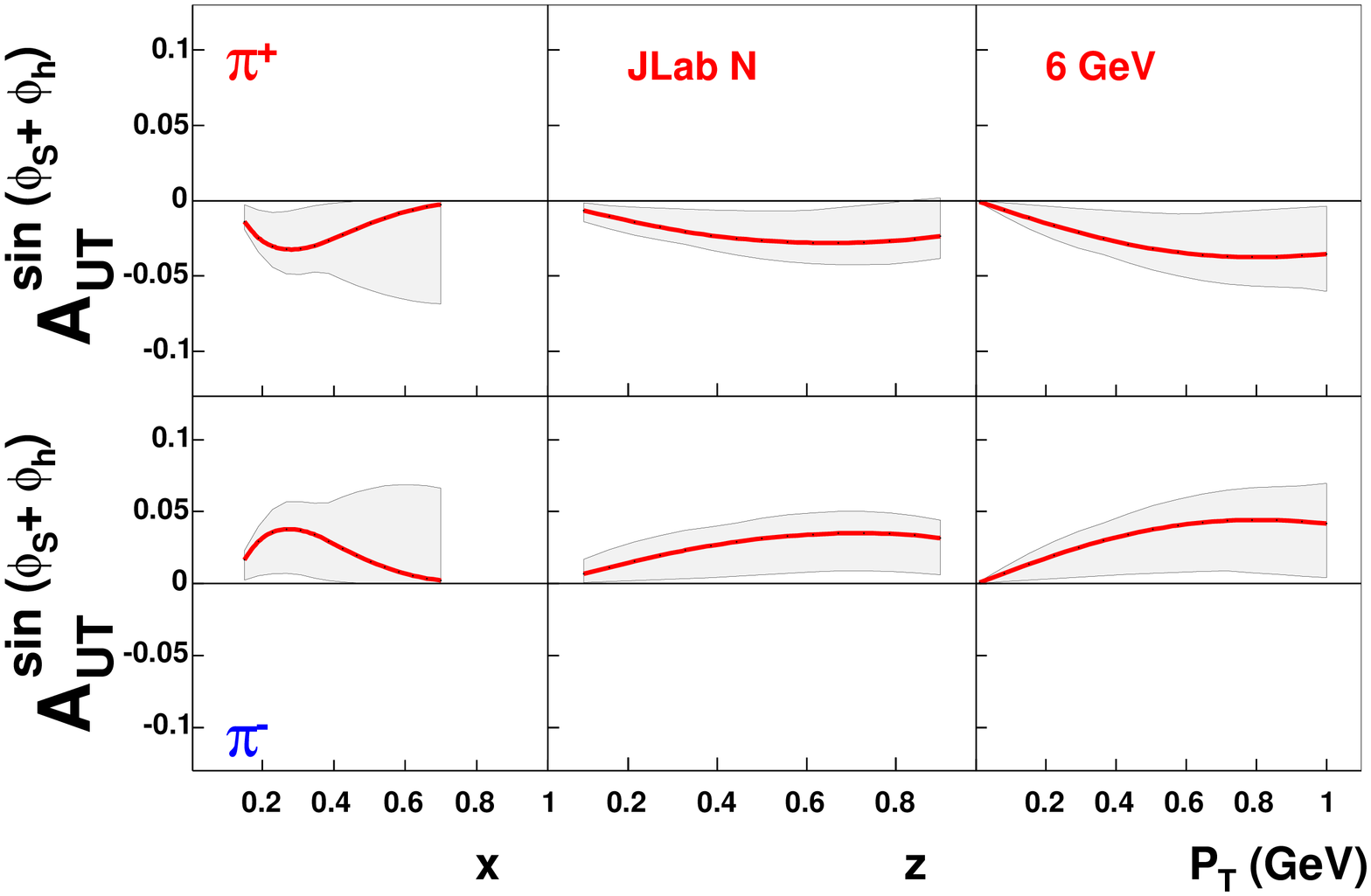}
\caption{\label{fig:jlab6}
Predictions for the single spin asymmetry $A_{_{UT}}^{\sin(\phi_S+\phi_h)}$ as
it will be measured at JLab operating on polarized hydrogen (proton, upper
plot) and He$^3$ (neutron, lower plot) targets at a beam energy of $6$ GeV.}
\end{figure}
%%%%%%%%%%%%%%%%%%%%%%%%%%%%%%%%%%%%%%%%%%%%%%%%%%%%%%%%%%%%%%%%%%%%%%%%%%%%%%%
%
%%%%%%%%%%%%%%%%%%%%%%%%%%%%%%%%%%%%%%%%%%%%%%%%%%%%%%%%%%%%%%%%%%%%%%%%%%%%%%%
\begin{figure}[t]
\vspace*{-0.7cm}
\includegraphics[width=0.33\textwidth,bb= 10 140 540 660,angle=-90]
{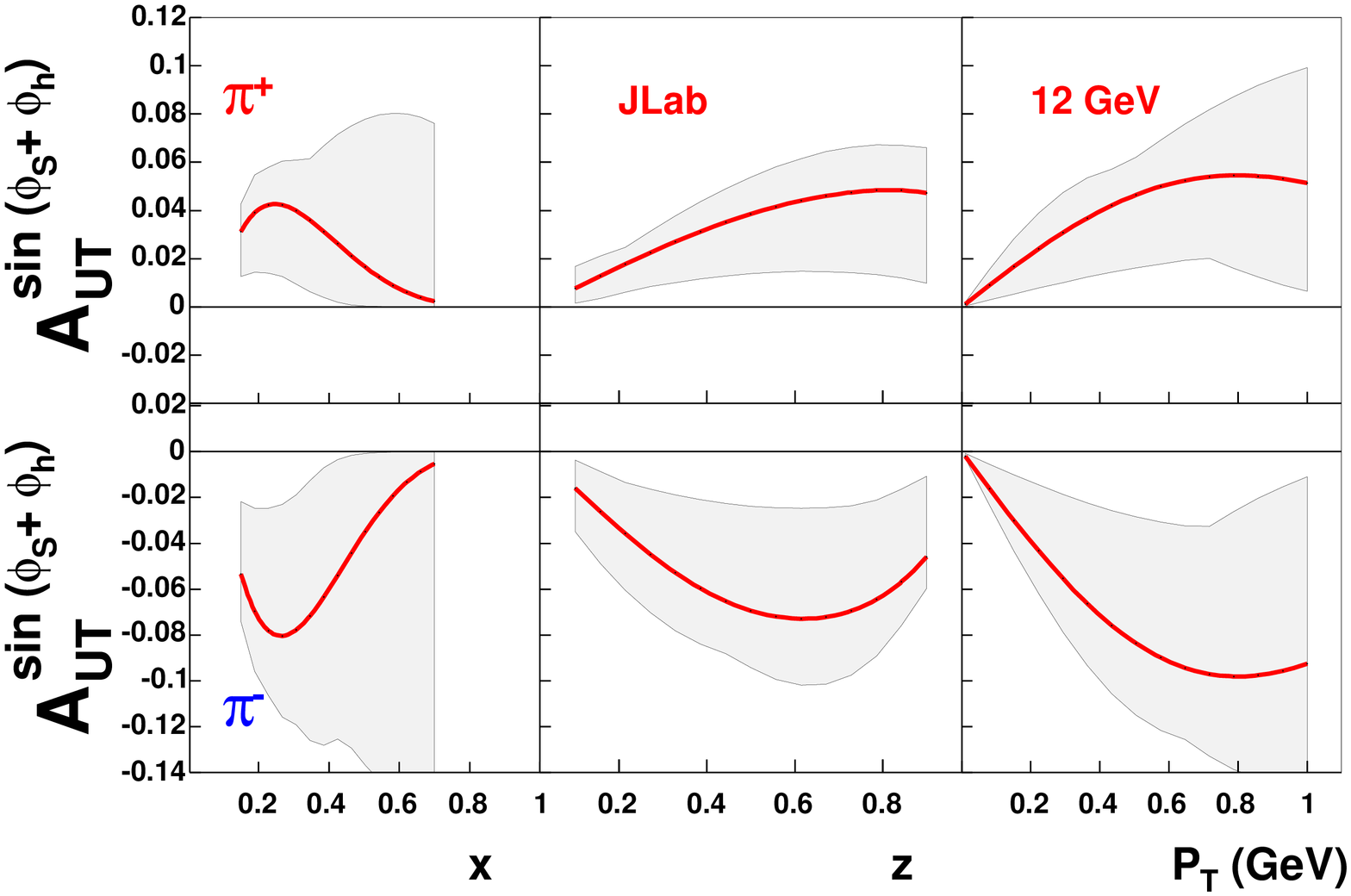} \\
\includegraphics[width=0.33\textwidth,bb= 10 140 540 660,angle=-90]
{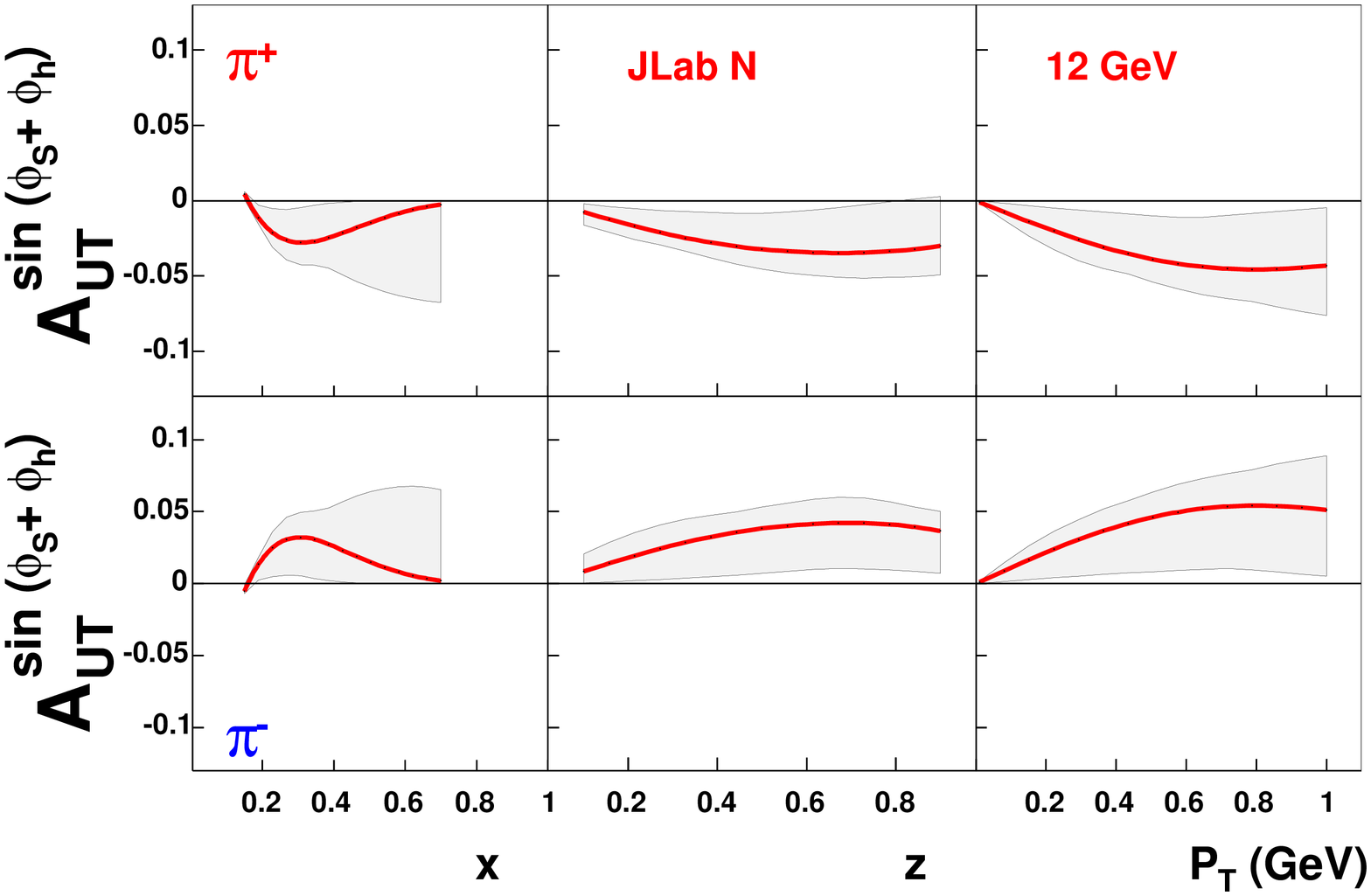}
\caption{\label{fig:jlab12}
Predictions for the single spin asymmetry $A_{_{UT}}^{\sin(\phi_S+\phi_h)}$ as
it will be measured at JLab operating on polarized hydrogen (proton, upper 
plot) and He$^3$ (neutron, lower plot) targets at a beam energy of $12$ GeV.}
\end{figure}
%%%%%%%%%%%%%%%%%%%%%%%%%%%%%%%%%%%%%%%%%%%%%%%%%%%%%%%%%%%%%%%%%%%%%%%%%%%%%%%

The JLab experiments will measure $A_{_{UT}}^{\sin(\phi_S+\phi_h)}$ for pion
production off transversely polarized proton and neutron targets, at incident
beam energies of either $6$ or $12$ GeV. The kinematical region spanned by
these experiments is very interesting, as it will enable to explore
the behavior of the transversity distribution function at large values of $x$,
up to $x \sim 0.6$. The adopted experimental cuts for JLab operating on a
proton target at $6$ GeV are the following
\be
\begin{tabular}{lll}
$0.4 \le z_h \le 0.7$, ~~~~~~ & $0.02 \le P_T \le 1 \;{\rm GeV}$,\\
$0.1 \le \xb \le 0.6$, ~~~~~~ & $0.4 \le y \le 0.85$, \\ 
$Q^2 \ge 1 \;{\rm GeV}^2$, ~~~~~~ & $ W^2 \ge 4\; {\rm GeV}^2$, \\ 
$1 \le E_h \le 4\; {\rm GeV} \;,$ ~~~~~~ & ~~~~~~
\end{tabular}
\label{JLab-6}
\ee
whereas for a beam energy of $12$ GeV they are
\be
\begin{tabular}{lll}
$0.4 \le z_h \le 0.7\;\;\;$, ~~~~~~ & $0.02 \le P_T \le 1.4 \;{\rm GeV}$, \\
$0.05 \le \xb \le 0.7$,  ~~~~~~ & $0.2 \le y \le 0.85$, \\ 
$Q^2 \ge 1 \;{\rm GeV}^2 $, ~~~~~~ & $W^2 \ge 4\;{\rm GeV}^2$, \\
$1 \le E_h \le 7\;  {\rm GeV}\;.$  & ~~~~~~ 
\end{tabular}
\label{JLab-12}
\ee
For a neutron target at $6$ GeV the cuts read:
\be
\begin{tabular}{lll}
$0.46 \le z_h \le 0.59$, & $0.13 \le \xb \le 0.40$, \\
$0.68 \le y \le 0.86$, & $1.3 \le Q^2 \le 3.1 \; {\rm GeV}^2 $, \\
$5.4 \le W^2 \le 9.3 \;  {\rm GeV}^2 $, ~~~~&
$2.385 \le E_h \le 2.404\;  {\rm GeV} \;,$
\end{tabular}
\label{JLab-neutron-6}
\ee
whereas for an incident beam energy of $12$ GeV they are:
\be
\begin{tabular}{lll}
$0.3 \le z_h \le 0.7$, ~~~~~~ &
$0.05 \le \xb \le 0.55$,\\ 
$0.34 \le y \le 0.9$,~~~~~~ & 
$Q^2 \ge 1 \;{\rm GeV}^2$, \\ 
$W^2 \ge 2.3\; {\rm GeV}^2 \;.$  ~~~~~~ & ~
\end{tabular}
\label{JLab-neutron-12}
\ee
Our corresponding predictions, according to Eq.~(\ref{sin-asym-final})
and our extracted transversity and Collins functions, are shown in
Figs.~\ref{fig:jlab6} and \ref{fig:jlab12}.

It is important to stress that, as the large $x$ region is not covered by the
HERMES and COMPASS experiments, our predictions for the $x$ dependence of
$A_{_{UT}}^{\sin(\phi_S+\phi_h)}$ are very sensitive to the few available
data points from HERMES and COMPASS at moderately large $x$ values.
As a consequence, the predictions for the JLab experiments may vary
drastically in the region $0.4 \le\xb\le 0.6$, as indicated by the large
shaded area in Figs.~\ref{fig:jlab6} and \ref{fig:jlab12}.
On the contrary, the results on the $P_T$ and $z_h$ dependences are
more stable, as they only depend on the transversity distribution
function integrated over~$x$.

\begin{figure}[h!]
\vspace*{-0.3cm}
\includegraphics[width=0.35\textwidth,bb= 10 140 540 660,angle=-90]
{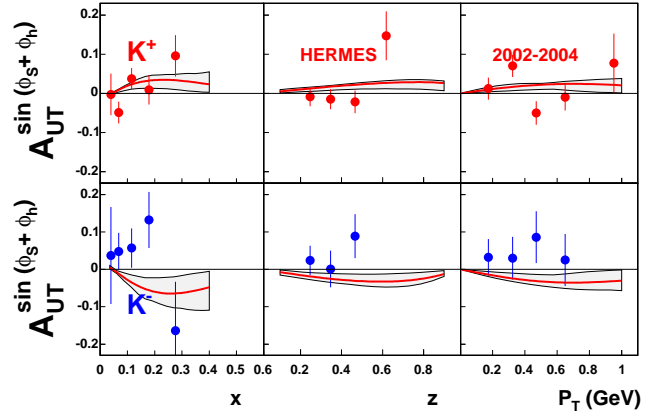}
\caption{\label{fig:hermeskaons}
Our results, based on the extracted transversity and Collins functions,
for the azimuthal asymmetry $A_{_{UT}}^{\sin(\phi_S+\phi_h)}$ for $K^\pm$
production, compared with the HERMES experimental data
\cite{Airapetian:2004tw,HERMES:proceedings}.}
\end{figure}

Finally, we compute the azimuthal asymmetry $A_{_{UT}}^{\sin(\phi_S+\phi_h)}$
for the production of $K$ mesons and compare it with existing HERMES
results~\cite{Airapetian:2004tw,HERMES:proceedings}. These data have not been included in our
best fit, as they might involve the transversity distribution of strange
quarks in the nucleon, which we have neglected for SIDIS data on $\pi$
production. We show our results in Fig.~\ref{fig:hermeskaons}, obtained using
the extracted $u$ and $d$ transversity distributions. Again, we have used
favored ($\Delta^ND_{K^+/u^\ua}$) and unfavored
($\Delta^ND_{K^-/u^\ua}, \Delta^ND_{K^\pm/d^\ua}$) Collins functions, as in
Eqs.~(\ref{coll-funct}), (\ref{NC}) and (\ref{h-funct}). For these we have
used the same parameters $N^{\C}_q, \gamma, \delta$ and $M$ of
Table~\ref{fitpar12}, with the appropriate unpolarized fragmentation functions
$D_{K^\pm/q}$~\cite{Kretzer:2000yf}.

We notice that our computations are in fair agreement with data concerning the
$K^+$ production, which is presumably dominated by $u$ quarks; instead, there
seem to be discrepancies for the $K^-$ asymmetry, for which the role of $s$
quarks might be relevant. New data on the azimuthal asymmetry
for $K$ production, possible from COMPASS and JLab experiments, might be
very helpful in sorting out the eventual importance of the sea quark
transversity distributions in a nucleon.

%
%%%%%%%%%%%%%%%%%%%%%%%%%%%%%%%%%%%%%%%%%%%%%%%%%%%%%%%%%%%%%%%%%%%%%%%%%%%%%%%
%%%%%%%%%%%%%%%%%%%%%%%%%%%%%%%%%%%%%%%%%%%%%%%%%%%%%%%%%%%%%%%%%%%%%%%%%%%%%%%

\section{\label{comm} Comments and Conclusions}

We have performed a combined analysis of all experimental data on spin
azimuthal asymmetries which involve the transversity distributions of
$u$ and $d$ quarks and the Collins fragmentation functions, classified
as favored (when the fragmenting quark is a valence quark for the final
hadron) and unfavored (when the fragmenting quark is not a valence quark
for the final hadron). We have fixed the total number of 9 parameters
by best fitting the HERMES, COMPASS and Belle data.

All data can be accurately described, leading to the extraction of the
favored and unfavored Collins functions, in agreement with similar results
previously obtained in the literature \cite{Vogelsang:2005cs,Efremov:2006qm}.
In addition, we have obtained, \textit{for the first time}, an extraction
of the so far unknown transversity distributions for $u$ and $d$ quarks,
$\Delta_T u(x)$ and $\Delta_T d(x)$. They turn out to be opposite in sign, with
$|\Delta_T d(x)|$ smaller than $|\Delta_T u(x)|$, and both smaller than their 
Soffer bound \cite{Soffer:1994ww}.

The knowledge of the transversity distributions and the Collins
fragmentation functions allows to compute the azimuthal asymmetry
$A_{_{UT}}^{\sin(\phi_S+\phi_h)}$ for any SIDIS process; we have then
presented several predictions for incoming measurements from COMPASS and
JLab experiments. They will provide further important tests of our complete
understanding of the partonic properties which are at the origin of SSA.
Data on $K$ production will help in disentangling the role of sea quarks.

Further expected data from Belle will allow to study in detail not only
the $z$ dependence of the Collins functions, but also their $\pp$ dependence.
The combination of data from SIDIS and $e^+e^- \to h_1 h_2 \, X$ processes
opens the way to a new phenomenological approach to the study of the nucleon
structure and of fundamental QCD properties, to be further pursued.

\begin{acknowledgments}
We are grateful to Werner Vogelsang for supplying us with the numerical
program for the QCD evolution of the transversity distributions.\\
We acknowledge the support of the European Community - Research Infrastructure 
Activity under the FP6 ``Structuring the European Research Area'' 
program (HadronPhysics, contract number RII3-CT-2004-506078).
\end{acknowledgments}

\bibliographystyle{h-physrev3.bst}
\bibliography{sample}

\end{document}